\documentclass[aps,prb,floatfix,epsf,epsfig,superscriptaddress]{revtex4}
\usepackage{graphicx}
\usepackage{graphics}
\usepackage{epsfig}
\usepackage{amsmath}
\usepackage{amssymb}
\newcommand{\BEQ}{\begin{equation}}
\newcommand{\EEQ}{\end{equation}}
\newcommand{\BEA}{\begin{eqnarray}}
\newcommand{\EEA}{\end{eqnarray}}

\renewcommand{\d}{{\rm d}}

\newcommand{\half}{\frac{1}{2}}

\newcommand{\F}{\widetilde{F}}
\newcommand{\x}{X}
\begin{document} 
\title
{Unzipping of DNA with correlated base-sequence.}
\date{\today}
\author{A.E. Allahverdyan}
\email{armena@science.uva.nl}
\affiliation{Institute for Theoretical Physics,
Valckenierstraat 65, 1018 XE Amsterdam, The Netherlands.}
\affiliation{Yerevan Physics Institute,
Alikhanian Brothers St. 2, Yerevan 375036, Armenia.}
\author{Zh.S. Gevorkian}
\email{gevorkia@phys.sinica.edu.tw}
\affiliation{Institute of Physics, Academia Sinica,
Nankang, Taipei 11529, Taiwan.}
\affiliation{Yerevan Physics Institute,
Alikhanian Brothers St. 2, Yerevan 375036, Armenia.}
\affiliation{Institute of Radiophysics and Electronics, Ashtarak-2 378410, 
Armenia.}
\author{Chin-Kun Hu}
\email{huck@phys.sinica.edu.tw}
\affiliation{Institute of Physics, Academia Sinica,
Nankang, Taipei 11529, Taiwan.}
\author{Ming-Chya Wu}
\email{mcwu@phys.sinica.edu.tw}
\affiliation{Institute of Physics, Academia Sinica,
Nankang, Taipei 11529, Taiwan.}

\begin{abstract}
  We consider force-induced unzipping transition for a heterogeneous
  DNA model with a correlated base-sequence. Both finite-range and
  long-range correlated situations are considered.  It is shown that
  finite-range correlations increase stability of DNA with respect to
  the external unzipping force.  Due to long-range correlations the
  number of unzipped base-pairs displays two widely different
  scenarios depending on the details of the base-sequence: either
  there is no unzipping phase-transition at all, or the transition is
  realized via a sequence of jumps with magnitude comparable to the
  size of the system.  Both scenarios are different from the behavior
  of the average number of unzipped base-pairs (non-self-averaging).
  The results can be relevant for explaining the biological purpose of
  correlated structures in DNA.

\end{abstract}
\pacs{
PACS: }
\maketitle


\section{Introduction.}

Structural transformations of DNA under changing of external
conditions are of importance for molecular biology \cite{bio} and
biophysics \cite{biofiz}.  They take place in transcription of genetic
information from DNA and in duplication of DNA during cell division
\cite{bio}.  The common scenario of these processes is unwinding of
the double-stranded structure of DNA under influence of external
forces.  Recall that a deoxyribonucleic acid (DNA) consists of two
strands with one winded around the other. These two strands interact via
hydrogen bonds due to which the double-helix structure is formed.  The
individual strand is constructed by covalent bonds whose strength is
thus much larger than the inter-strand coupling. Each strand is a
polymer based on nucleotides.  A nucleotide is a deoxyribose sugar
molecule bearing on one side purine or pyrimidine group (the base) and
on the other a phosphate group. The purines can be of two type:
adenine (A) and guanine (G), whereas pyrimidines are cytosine (C) and
thymine (T) (an additional purine uracil (U) is found in ribonucleic
acid (RNA)).  A, G, C and T groups differentiate the nucleotides and
constitute the genetic code carried by a DNA molecule. The bounds
between neighboring nucleotides within one strand are formed via the
corresponding phosphate groups. Hydrogen bonds between opposite
strands are formed either by A-T bases or by G-C bases.  Since the
bases A, G, C and T are hydrophobic, they are located at the core of
the double-helix. In contrast, the sugar molecules and the phosphate
groups are hydrophilic and they are located in the outside part of the
DNA molecule. Thus in a regular DNA molecule the letters of the
genetic code are hidden from the molecular environment. This appears
as a problem for the polymerase enzymes whose role is to read the
genetic code. The polymerase may function if only they unzip the
needed part of the DNA molecule, so that the bases are exposed to the
environment. This is the main reason why DNA unzipping, in particular,
unzipping under an external force is important for functioning of all
living organisms.  Force-induced unzipping has been actively
investigated only recently ~\cite{hindu,non-hindu,maritan,nelson,exp}
motivated by the new generation of micromanipulation experiments~
\cite{exp,expreview}.

It is expected that features of the unzipping process depend on the
base-sequence of DNA, because AT and GC base-pairs do have different
formation energies. It is more difficult to break a single GC
base-pair, since it is made of three hydrogen bonds, while a single AT
base-pair is made of two hydrogen bonds only. Thus, the formation
energy difference between AT and GC base-pairs is of the order of one
hydrogen bond energy, that is, $0.1-0.2$ eV. This is comparable with
the average formation energy itself. We note in addition that for
a given DNA molecule the overall concentrations of AT and GC
base-pairs are approximately equal \cite{bio}.  This is especially
true for higher organisms, e.g., the concentration of GC base-pairs
for primates is between 49 and 51\% \cite{bio}.

The above energy difference may not be relevant for certain bulk
properties of DNA.  Therefore, the latter is frequently modeled
assuming a homogeneous base-sequence.  However, in natural conditions
the energy supplied for uzipping can be comparable to the average
formation energy, and then the heterogeneous character of the
base-sequence becomes relevant.  One of the first steps in this direction
was made in \cite{nelson}, where it was shown that short-range
heterogenity does influence the unzipping process in the region where
the energy supplied by an external unzipping force is comparable to the
average formation energy of a DNA base-pair.

Our main purpose is to make the next step towards real DNAs and to
analyze force-induced unzipping for a DNA-model, where the structural
features of the base-sequence are taken into account.  One of the
known features of DNA is that its base-sequence displays substantial
correlations which, in particular, can be of long-range characters
\cite{longrange,voss,coding,review}: two base pairs separated from
each other by thousands of pairs appear to be statistically
correlated.  Initial studies reported long-range correlations for
non-coding regions of DNA (introns).  For higher organisms, e.g.
humans, these regions constitute more than 90\% of DNA \cite{bio}.  It
was believed for some time that coding regions, which carry the
majority of genetic information, can have only short-range
correlations.  However, more recent results indicate on the existence
of weak long-range correlations in coding regions as well
\cite{coding} (this point was controversial for a while, but the
general consensus on its validity emerged gradually).  Moreover,
systematic changes were found in the structure of correlations
depending on the evolutionary category of the DNA carrier \cite{voss}.
In spite of ubiquity of long-range correlations, their biological
reason remains largely unexplored. Some attempts in this direction
were made in ~\cite{jenya}, where it was studied why long-range
correlations are absent in certain biologically active proteins.

Our basic purpose in the present paper will be to determine how
statistical correlations, in particular long-range correlations,
influence on the unzipping process. Due to the biological relevance of
unzipping, indications of such influences can provide useful
information for explaining the presence of long-range correlations in
DNA.

This paper is organized as follows. The basic model we work with is
described in section \ref{model}. The situation with finite-range
correlated base-sequence is investigated in section \ref{OUnoise}. The next 
three sections study various aspects of the long-range correlated
situation. We conclude with a summary of our results. Several technical
points are outlined in appendices.

\section{The model} 
\label{model}

There are three basic mechanisms which determine the physics
of the unzipping process: An external force tending to unbind the
double-helix structure of a DNA molecule, thermal noise generated by
an equilibrium environment into which the molecule is embedded, and
finally structural features of the molecule itself. Among various
structural features which may be of relevance, the most important ones
are connected with the base-sequence of the molecule.

We shall work with a model which takes into account these three
physical ingredients in the most minimal way. It was recently proposed
in Ref.~\cite{nelson} for studying DNA unzipping.

{\it i)} A DNA molecule is lying along the $x$-axis between
the points $x=a$ and $x=L$. 

{\it ii)} Among all degrees of freedom of the molecule we consider only
base-pairs; they are located at points $x_i$, $a<x_i<L$,
$i=1,...,M$. Indeed, for that range of external force where the
molecule is close to be unbind completely, those degrees of freedom
which are related to hydrogen bonds have much shorter characteristic
times as compared to other degrees of freedom. The latter ones can
therefore be considered as adiabatically frozen, and excluded from the
effective description we are developing.

{\it iii)} Any base-pair can be in one of two states: bound or
disconnected (broken). We choose the overall energy scale in such a
way that the latter case contributes to the Hamiltonian a binding
energy $\phi(x_i)$, whereas the former case brings nothing.  
As we stressed in the introduction,
different
types of base-pairs do have different binding energies: even when
considering the ideal situation, where there are no ``wrong
base-pairs'' such as AC and GT, the ``correct'' base-pairs AT and
GC are different with respect to energy needed to unbind them.  Thus
$\phi(x_i)$ is a random quantity with an average $\langle\phi\rangle$:
\BEA \phi(x_i)=\langle\phi\rangle+\eta(x_i).  \EEA

{\it iv)} An external force is acting
on the left end $x=a$ of the molecule pulling apart the two strands.
Thus, if a bond $x_i$ is broken, all the base-pairs $x_j$ with $j<i$ are
broken as well. Each broken bond brings additionally to the
Hamiltonian a term $-{\cal F}$, where ${\cal F}$ is proportional to
the acting force. 

{\it v)} Summarizing all of these, one comes to the
Hamiltonian 
\BEA
H(x)=-{\cal F}x+\sum_{i=1}^x
\phi(x_i)=(\langle\phi\rangle-{\cal F})x+\sum_{i=1}^x \eta(x_i),
\EEA
where $x$ is the number of broken base-pairs. 

In the thermodynamical limit,
where $L,M\gg 1$, one applies the continiuum description with $x$
being a real number, $a<x<L$, and ends up with the following
Hamiltonian:
\BEA
\label{d1}
H(x)=(x-a)\,f+\int_a^x\d s\,\eta(s),
\EEA
where $f=\langle \phi\rangle-{\cal F}$ and $\beta=1/T$ is the inverse
temperature ($k_{\rm B}=1$). 

{\it vi)} For characteristic time-scales of unzipping experiments we
can certainly neglect any changes of the base-sequence for a single
DNA molecule. Thus, once it is modeled via the random noise $\eta$,
it is legitimate to assume that this noise is frozen, i.e. its single
realization corresponds to a single molecule. It is assumed that the
DNA molecule is embedded into a thermal bath with temperature $T$, and
had sufficient time to reach equilibrium.  Thus, the partition
function and the free energy corresponding to the Hamiltonian
(\ref{d1}) read:
\BEA
\label{333}
Z=\int_a^L\d x\, e^{-\beta H(x)},
\qquad F=-T\ln Z.
\EEA
These quantities are still random together with $\eta$. Average results
of many experiments with various realizations of $\eta$ can be described
with help of the average free energy $\langle F\rangle$.
Our order parameter is the number of
broken base-pairs $\x$. Along with its average it is defined 
for $t=0$ as 
\BEA
\label{brams}
\x=\partial_fF,\quad
\langle \x\rangle=\partial_f\langle F\rangle.
\EEA

\subsection{Finite-range and long-range correlated situations.}

{\it vi)} It remains to specify the properties of the noise $\eta$.
Within the adopted description we assume it is a Gaussian stationary
process with an autocorrelation function
\BEA
\label{mimi}
K(t-t')=\langle\eta(t)\eta(t')\rangle,\qquad K(t)=K(-t).
\EEA

Two major classes can now be distinguished depending on the behavior of
$K(t)$ for large $t$.
The finite-range correlated situation is defined by requiring that the
integral 
\BEA
\label{in}
D=\int_0^\infty\d s\,K(s),
\EEA
determining the total intensity of the noise is finite. There are 
three particular case of the finite-range correlated situation.
The white noise case, 
\BEA
\label{delta}
K(t)=D\delta(t),
\EEA
describes completely uncorrelated noise. The physical situation given
by (\ref{d1}, \ref{delta}) is well known, and was used to describe
interfaces, random walks in a disordered media, and population
dynamics \cite{manfred}.  It was recently applied for the unzipping
transition in DNA \cite{nelson}. Similar models were considered in
\cite{maritan,azbel}.  

The second case corresponds to the noise having some finite
| though possibly large | correlation length $\tau$. The simplest and
most widely used model for this case is provided by 
Ornstein-Uhlenbeck (OU) noise 
\BEA
K(t)=\frac{D}{\tau}\,e^{-|t|/\tau},
\label{OU}
\EEA
where $D$ is the total intensity of the noise, 
and $\tau$ is the correlation time; 
$\tau\to 0$ corresponds to the white noise. 
The third case is when $K(t)$ has a power-law dependence for large
$t$, but still decays sufficiently quickly so that the integral in 
(\ref{in}) is finite: $K(t)\propto |t|^{-\delta}$ with $\delta>1$.

The second major class is the long-range correlated situation, where
the integral in (\ref{in}) is infinite, that is when $K(t)$ for
sufficiently large $t$ behaves according to a power law \cite{voss}:
\BEA
\label{korund}
K(t)\equiv\langle\eta(t)\eta(0)\rangle =\sigma |t|^{-\alpha},
\EEA 
where 
\BEA
0<\alpha<1 
\EEA
is the exponent characterizing the long-range
correlation, and where $\sigma$ is the (local) intensity. 
Note that $K(t)$ has to be regular and finite for small $t$ 
\cite{review}, as one would expect from physical reasons.

The OU noise (\ref{OU}), as the typical representative of the
finite-range correlated situation, and the long-range correlated noise
(\ref{korund}) are relevant for modeling correlations in
base-sequence of DNA \cite{longrange,voss,coding,review,maria}.  Note,
however, that the real noise distributions in DNA can be much more
complicated \cite{voss,coding}.  In particular, this concerns the
Gaussian property we assume (see in this context section \ref{un},
where we study a model of a non-Gaussian noise to show that its
predictions in the thermodynamic limit do not differ from those given
by the corresponding Gaussian noise). For the long range correlated
situation there can exist several characteristic exponents for
different ranges of $t$.  Nevertheless, Eqs.~(\ref{OU}, \ref{korund})
are certainly the minimal models of noise which are sufficiently
simple and which allow to study both finite and long range
correlations.

\subsection{Reduction to Langevin equation.}

The basic method of solving the present model will be to reduce it to the
physics of a Brownian particle whose dynamics is described by a 
stochastic differential equation. 
In Eq.~(\ref{333}) one fixes $L$, and views $a$ as a parameter varying
from the highest possible value $L$, where $Z=0$, to the lowest
possible value which we define to be $a=0$. The quantity $t=-a$ will
thus monotonicaly increase and can be interpreted as a time-variable.
Differentiating $Z$ in (\ref{333}) over $a$ and changing the variable as
$t=-a$, one gets:
\BEA
\label{d2}
\frac{\d Z}{\d t}=1-\beta f Z-\beta\,\eta(t)Z,
\qquad -L<t<0
\EEA
where we used $\eta(t)=\eta(-t)$, as follows from the Gaussian stationary
property of the noise. This is a Langevin equation with
a multiplicative noise. From (\ref{d2}) one can obtain a stochastic equation 
for $F=-T\ln Z$:
\BEA
\label{d3}
\frac{\d F}{\d t}+V'(F)=\eta(t),\quad
V(F)=T^2e^{\beta F}-fF.
\EEA
This is the basic stochastic equation we will work with.

\section{ Finite range correlated noise.} 

\subsection{Ornstein-Uhlenbeck noise.}
\label{OUnoise}

Our main purpose here is to study the process of
unzipping in the presence of the finite range corelated noise given by
Eq.~(\ref{OU}).  We wish to understand how the magnitude of $\tau$
influences unzipping.

Note that the OU noise (\ref{OU}) can itself be modeled via a white-noise:
\BEA
\tau\,\dot{\eta}=-\eta+\sqrt{D}\,\xi(t), 
\label{b1}
\EEA
where $\xi(t)$ is a Gaussian noise with delta-correlated spectrum: 
\BEA
\label{b2}
\langle\,\xi(t)\xi(t')\,\rangle=2\delta(t-t').
\EEA
Indeed, Eq.~(\ref{OU}) is recovered directly from (\ref{b1}, \ref{b2}),
since their exact solution is:
\BEA
\langle\,\eta(t)\eta(t')\,\rangle=
e^{-(t+t')/\tau}\left( \langle\,\eta^2(0)\,\rangle-\frac{D}{\tau}\right)
+\frac{D}{\tau}\,e^{-|t-t'|/\tau}.
\EEA
We get back from here Eq.~(\ref{OU}) under an additional consistency
condition $\langle\,\eta^2(0)\,\rangle={D}/{\tau}$.
Moreover, $\eta(t)$ is Gaussian random process, because $\xi(t)$
is Gaussian and Eq.~(\ref{b1}) is linear \cite{risken}.

To handle (\ref{d3}) one differentiates it over $t$ and uses 
(\ref{b1}, \ref{b2}).
Changing the variable as $s=t/\sqrt{\tau}$ one gets \cite{hanggi}:
\BEA
\label{bukhara}
\frac{\d ^2 F}{\d s^2}+\gamma(F)\,
\frac{\d F}{\d s}=-V'(F)+\frac{\sqrt{D}}{\tau^{1/4}}\xi(s),
\EEA
where 
\BEA
\gamma(F)=\tau^{-1/2}+V''(F)\,\tau^{1/2}.
\EEA  
Eq.~(\ref{bukhara}) has the same form as a Langevin equation for a
particle with unit mass in the potential $V(F)$ and subjected to a
white noise and a $F$-dependent friction with a coeffcient
$\gamma(F)$. Note that the potential $V(F)$ is confining only for
$f>0$: $V(F)\to \infty$ for $F\to\pm\infty$.

We can rewrite Eq.~(\ref{bukhara}) introducing an additional variable
$\F(s)={\d F(s)}/{\d s}$,
which in the above language of the Brownian motion
corresponds to the velocity.
\BEA
\label{b3}
&&\frac{\d F}{\d s}=\F,\\
&&\frac{\d \F}{\d s}=-\gamma(F)\,\F
-V'(F)+\frac{\sqrt{D}}{\tau^{1/4}}\xi(s).
\label{b4}
\EEA
As $\xi(t)$ is a Gaussian white noise, one uses the standard tools, see
e.g. \cite{risken}, and writes down from (\ref{b3}, \ref{b4})
a Fokker-Planck-Klein-Kramers
equation for the common probability distribution:
\BEA
P(F,\F,s)=\langle\,
\delta(F-F(s))\,\delta(\F-\F(s))
\,\rangle,
\EEA
where $F(s)$ and $\F(s)$ are particular noise-dependent solutions
of (\ref{b3}, \ref{b4}), and where the average is taken over 
the white noise $\xi(t)$
given by Eq.~(\ref{b2}).
\BEA
\label{fpkk}
\frac{\partial P(F,\F,s)}{\partial s}=-\F\,
\frac{\partial P(F,\F,s)}{\partial F}+\gamma(F)\,
\frac{\partial[\, \F\,P(F,\F,s)\,]}{\partial \F}+V'(F)\,
\frac{\partial P(F,\F,s)}{\partial \F}+\frac{D}{\tau^{1/2}}\,
\frac{\partial P(F,\F,s)}{\partial \F^2}.
\EEA
Our interest is in the large-$s$ limit of this equation (thermodynamic
limit), and we want to have the reduced probability distribution
$P(F,s)$ of $F$ only:
\BEA
P(F,s)=\langle\,\delta(F-F(s))\,\rangle=\int \d\F\,P(F,\F,s).
\label{agrav}
\EEA
To this end let us introduce
\BEA
Q_n(F,s)=\int \d\F\,\F^n\,P(F,\F,s)\qquad n=0,1,2,3..,
\EEA
where $Q_0(s)=P(F,s)$. From (\ref{fpkk}) one gets an infinite set
of coupled equations for $Q_n(F,s)$:
\BEA
\label{c1}
&&\frac{\partial Q_0(F,s)}{\partial s}=-
\frac{\partial Q_1(F,s)}{\partial F},\\
\label{c2}
&&\frac{\partial Q_1(F,s)}{\partial s}=-
\frac{\partial Q_2(F,s)}{\partial F}-\gamma(F)\,Q_1(F,s)-V'(F)\,Q_0(F,s),\\
\label{c3}
&&\frac{\partial Q_2(F,s)}{\partial s}=-
\frac{\partial Q_3(F,s)}{\partial F}-2V'(F)\,Q_1(F,s)
-2\gamma(F)\,Q_2(F,s)+\frac{2D}{\tau^{1/2}}\,Q_0(F,s),\\
&& ......
\EEA
When deriving (\ref{c1}--\ref{c2}) we used integration by parts, and
the following standard boundary conditions:
\BEA
P(F,\F,s)\to 0,\quad {\rm if}\quad F\to\pm\infty,
\quad {\rm or}\qquad {\rm if}\quad \F\to\pm\infty.
\label{osa1}
\EEA
These conditions are physically meaningful if the potential $V(F)$
is confining, and thus the motion of the corresponding Brownian particle
takes place in a finite domain.
According to the above discussion on the confining character of the
potential $V(F)=T^2e^{\beta F}-fF$, the boundary conditions 
(\ref{osa1}) are reliable only for $f>0$.

Recall that the ``time-variable'' $t$ moves between $-|L|$ and $0$.
For large lengths, i.e. for $L\gg 1$ (thermodynamic limit) and as the
consequence $t\propto s\to 0$, any solution of the equation
(\ref{fpkk}) relaxes towards the unique stationary distribution
$P_{\rm st}(F,\F)$. A rather general proof of this fact is presented
in \cite{risken}.

We shall now use (\ref{c1}--\ref{c2}) to get explicitly the stationary
distribution function $P_{\rm st}(F)$ of $F$. Putting to zero the LHS
of (\ref{c1}) one gets that $Q_{1,{\rm st}}(F)$ does not depend on
$F$. Taking into account the boundary condition (\ref{osa1}) one concludes
that it is equal to zero: 
\BEA
\label{c7}
Q_{1,{\rm st}}(F)=0.
\EEA
Putting to zero the LHS of (\ref{c2}) and using (\ref{c7}) we get
\BEA
\label{c8}
\frac{\partial Q_{2,{\rm st}}(F)}{\partial F}=-V'(F)\,
Q_{0,{\rm st}}(F).
\EEA
It remains to determine $Q_{2,{\rm st}}(F)$ putting to zero
the LHS of Eq.~(\ref{c3}).
One can conjecture that the stationary state $P_{\rm st}(F,\F)$ 
is symmetric with respect to $\F\to -\F$, and then $Q_{3,{\rm st}}(F)
=0$ in the same way as for $Q_{1,{\rm st}}(F)$ in (\ref{c7}).
Alternatively, one can assume that $\gamma(F)$ and $D$ are sufficiently
large so that the term $\partial Q_{3,{\rm st}}(F)/\partial F$
can be simply dropped in the RHS of (\ref{c3}). If
$V''(F)$ is of order one, then a large $\gamma(F)$ is realized both
for large and small $\tau$ \cite{hanggi}. Thus we conclude from
(\ref{c3}):
\BEA
\label{c9}
\gamma(F)\,Q_{2,{\rm st}}(F)=\frac{D}{\tau^{1/2}}\,Q_{0,{\rm st}}(F).
\EEA
In view of (\ref{c8}, \ref{c9}) one has a single differential equation:
\BEA
\frac{D}{\sqrt{\tau}}\,\frac{\partial }{\partial F}
\left(\frac{Q_{0,{\rm st}}(F)}{\gamma(F)}\right)=-V'(F)\,
Q_{0,{\rm st}}(F),
\EEA
and gets for $Q_{0,{\rm st}}(F)\equiv  P_{\rm st}(F)$,
\BEA
P_{\rm st}(F)&&\propto\,\gamma(F)\exp\left[-\frac{\tau}{2D}[V'(F)]^2
-\frac{1}{D}V(F)\right],\nonumber\\
P_{\rm st}(F)&&=\,{\cal N}(1+\tau\,e^{\beta F})
\exp\left[\frac{fF}{D}-
\frac{T\left(T-\tau f\right)e^{\beta F}}{D}
-\frac{\tau T^2}{2D}\,e^{2\beta F}\,\right].
\label{trick}
\EEA
where ${\cal N}$ is the normalization factor. 
The white-noise, $\tau\to 0$, limit of $P_{\rm st}(F)$ was obtained in
Refs.~\cite{manfred,nelson}. 

\begin{figure}[bhb]
\includegraphics[width=10cm]{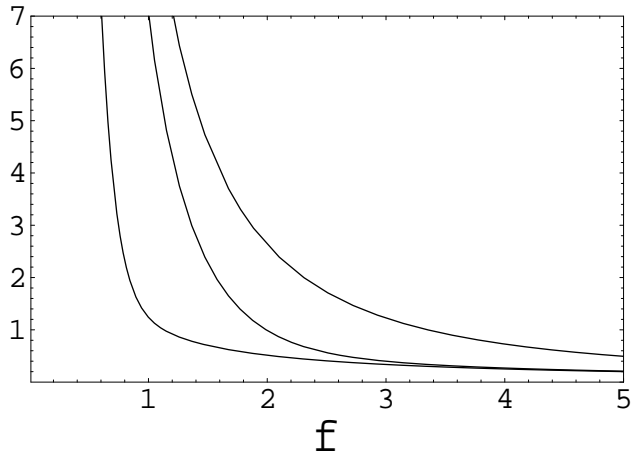}
\caption{$\langle \x\rangle$ 
for Ornstein-Uhlenbeck noise with $D=10$, $T=1$. 
From right to left: $\tau=0$, 
$\tau=10$, $\tau=100$. It is seen that, for a fixed $f$,
$\langle \x\rangle$ decreases upon increasing $\tau$.}
\label{f1}
\end{figure}

According to (\ref{brams}, \ref{trick}) the average free energy reads:
\BEA
\langle F\rangle=T\,\frac{\int_0^\infty\d u\,\ln (u)
\left(\tau+\frac{1}{u}\right)\,u^{\mu}
\,\exp\left[
\left(\mu\tau-\frac{T^2}{D}\right)\,u
-\frac{\tau T^2}{2D}\,u^2\right]}
{\int_0^\infty\d u
\left(\tau+\frac{1}{u}\right)\,u^{\mu}
\,\exp\left[
\left(\mu\tau-\frac{T^2}{D}\right)\,u
-\frac{\tau T^2}{2D}\,u^2\right]},
\label{wolf1}
\EEA
where
\BEA
\mu=\frac{Tf}{D}.
\EEA

Note that both integrals in (\ref{wolf1}) can be expressed
through the gamma-function $\Gamma(x)$ 
and the confluent hypergeometric (Kummer) function $_1F_1(a,b;z)$, since
\BEA
\int_0^\infty\d u\, u^c\,e^{au^2-bu}=
\frac{1}{2}\,a^{-1-c/2}\,\Gamma\left(
1+\frac{c}{2}\right)\left[
b~_1F_1\left(1+\frac{c}{2},\frac{3}{2},\frac{b^2}{4a}\right)
+\sqrt{a}~_1F_1\left(\frac{1+c}{2},\frac{1}{2},\frac{b^2}{4a}\right)
\right].
\EEA
Similar formulas can be written for 
$\int_0^\infty\d u\, u^c\,(\,\ln u\,)^n\,e^{au^2-bu}$ for $n=1,2$.
These representations facilitate numerical calculations.

The average number of broken base-pairs 
$\langle \x\rangle$ can be calculated from (\ref{brams}, 
\ref{trick}). Note that for the white-noise situation $\tau\to 0$
a simple formula is obtained:
\BEA
\langle \x\rangle=\frac{T^2}{D}\,\frac{\d \psi(\mu)}{\d \mu},
\EEA
where $\psi(\mu)=\Gamma(\mu)/\Gamma '(\mu)$. For $\mu\to 0$, $\langle \x
\rangle$
does not depend on temperature and on $\tau$ and becomes very large:
\BEA
\label{tru}
\langle \x\rangle
=D\,f^{-2},
\EEA
for $f\to 0$. When the external force reaches its critical value,
the average number of broken base-pairs diverges in the thermodynamic limit.

To study the influence of $\tau$ on this unzipping phase transition,
one should keep in mind the realistic situation, where DNA molecules
belonging to different evolutionary classes have different correlation
properties of their base-sequences \cite{voss}. At the same time the
concentration (fraction) of AT and GC base-pairs is known to be
(approximately) equal for sufficiently long DNA molecules in natural
conditions \cite{bio,f1}.  Therefore, in comparing two situations
having different correlation characteristics, it is legitimate to keep
fixed the intensity of the noise defined by Eq.~(\ref{in}) | 
this corresponds to fixed concentration of various base-pairs | 
and to study how the average number of broken base-pairs $\langle
\x\rangle$ depends on $\tau$ for some fixed value of $f$. This 
dependence is displayed in
Fig.~(\ref{f1}) following to Eqs.~(\ref{brams}, \ref{wolf1}).  It is
seen that the behavior of $\langle \x\rangle$ for very small $f$ 
depends on $\tau$ rather weakly.  Indeed, as follows from Eq.~(\ref{trick}),
for $f\to 0$ the relevant domain of $F$ contributing into $\langle
F\rangle$ is $F\sim -D/f$. As it does not depend on $\tau$, we get
back Eq.~(\ref{tru}).  However, a non-trivial
dependence on $\tau$ does exist for moderately small values of $f$,
where as seen in Fig.~\ref{f1}, $\langle \x\rangle$ is a decreasing
function of $\tau$ for a fixed $f$: longer correlations present in the
base-sequence increase the stability of the DNA molecule, since larger
external forces ${\cal F}$ needed to achieve the same average amount
of broken base-pairs.  This is our main qualitative conclusion on how
a finite correlation length influences the unzipping process.

\subsection{Arbitrary finite-range correlated noise at low temperatures.}
\label{lowT}

In the previous section we reduced the non-linear equation (\ref{d3})
with the finite range correlated noise (\ref{OU}) to a Fokker-Planck
equation, and solved the latter exactly in the thermodynamic limit.
The essential feature that made this analytical solution possible is
that the OU noise has a single and well-defined characteristic time
and due to this allows representation (\ref{b1}, \ref{b2}).

In general it is impossible to solve (\ref{d3}) for an arbitrary
Gaussian noise, and, in particular, for the situation given by
Eq.~(\ref{korund}): there is no exact Fokker-Planck equation for this
case.  There is, however, a particular case which allows analytical
treatment.  For very low temperatures, $T\to 0$, one can approximately
substitute $V(F)$ in Eq.~(\ref{d3}) by $-fF$ for $F<0$ and by an
infinite potential wall standing at $F=0$. Thus, all values $F>0$
become prohibited.  For this particular form of potential one can get
a Fokker-Planck equation for 
\BEA
P(F,t)=\langle\delta(F-F[\eta,t])\rangle
\label{set}
\EEA
with an {\it arbitrary} Gaussian noise in the RHS of Eq.~(\ref{d3}).
The derivation goes as follows. Write Eq.~(\ref{d3}) as
\BEA
\label{kato}
\frac{\d F}{\d t}=f+\eta(t),
\EEA
where the stochastic variable $F$ is restricted to be negative due
to the above infinite wall. Differentiating $P(F,t)$ in (\ref{set})
over $t$ one gets
\BEA
\label{ko1}
\frac{\partial P(F,t)}{\partial t}
=-f\,\frac{\partial P(F,t)}{\partial F}
-\frac{\partial }{\partial F}\,\langle \eta(t)\,\delta(F-F[\eta,t])\rangle.
\EEA
It remains to handle the last term in this equation. One uses 
Novikov's theorem \cite{fox}
\BEA
\label{novi}
\langle\eta(t)\,\delta(F-F[\eta,t])\,\rangle
=-\frac{\partial}{\partial F}
\int_{-L}^t\d s\,K(t-s)\left\langle\,
\delta(F-F[\eta,t])\,\frac{\delta
F[t]}{\delta\eta(s)}\,\right\rangle,
\EEA
where $\delta \,/\delta\eta(s)$ is the variational 
derivative and $\delta F[t]/\delta\eta(s)$ is obtained 
from (\ref{kato}):
\BEA
\label{koko1}
\frac{\d }{\d t}\frac{\delta F[t]}{\delta\eta(s)}
=\delta(t-s),\qquad \frac{\delta F[t]}{\delta\eta(s)}=\theta(t-s),
\EEA
where $\theta(t-s)$ is the step function.
Combining (\ref{ko1}, \ref{novi}, \ref{koko1}) we get finally
\BEA 
\frac{\partial P(F,t)}{\partial t} =-f\,\frac{\partial P(F,t)}{\partial F}
+D_t\frac{\partial ^2P(F,t)}{\partial F^2},
\label{do}
\EEA
\BEA
D_t=\int_0^{t+L}\d s\,K(s).
\label{dodosh}
\EEA
Eq.~(\ref{do}) should additionally be supplemented by a boundary
condition which reflects the presence of the infinite wall at
$F=0$. Equation~(\ref{do}) can be written as the continuity equation
\BEA
\frac{\partial P(F,t)}{\partial t} +\frac{\partial J(F,t)}{\partial F}=0,\qquad
J(F,t)=f-\mu_t\frac{\partial P(F,t)}{\partial F},
\EEA
where $J(F,t)$ is the probability current. The infinite wall at $F=0$
is now implemented by requiring:
\BEA
\label{kaa1}
&& \int_{-\infty}^0\d F\,P(F,t)=1,\\
&& J(0,t)=0,
\label{kaa2}
\EEA
for all $t$. Conditions (\ref{kaa1}, \ref{kaa2}) are imposed on any
solution of (\ref{do}).

In the thermodynamic limit $L\to \infty$ and $t=0$ 
one gets from the stationarity condition ${\partial P(F,t)}/{\partial t}=0$
\BEA
\label{no1}
P(F)=&&\frac{f}{D}\,
         \exp\left[\frac{fF}{D}\right], \qquad {\rm for}\qquad F<0,\\ 
      =&& 0, \qquad\qquad\qquad\quad {\rm for}\qquad F\geq 0,
\label{no2}
\EEA
where the total intensity, as given by (\ref{in}), is finite for the
considered short-range correlated situation.  Note that in the
thermodynamical limit conditions (\ref{kaa1}, \ref{kaa2}) are
satisfied automatically as seen from Eqs.~(\ref{no1}, \ref{no2}).  It
is now seen from (\ref{brams}) that
\BEA
\label{drastamat}
\langle \x\rangle=D f^{-2}, 
\EEA
which has the same $f$-dependence as the white-noise case for small
$f$; see (\ref{tru}). We conclude that, not unexpectedly, for low
temperatures the behavior of $\x$ is determined
only the total intensity of the noise. All other details of $K(t)$ 
do not matter. It remains to stress that the present analysis certainly
does not apply to the long-range correlated situation (\ref{korund}),
since the total intensity $D$ diverges in the thermodynamical limit.

In closing this section, let us note that Eq.~(\ref{drastamat})
can be applied to finite-range correlated noise that for 
$t\ll L$ has the same autocorrelation function as (\ref{korund}).
As an example take
\BEA 
\label{srika}
K_{\rm fr}(t)=&&\sigma |t|^{-\alpha},\quad {\rm for}\quad |t|\leq l,\\
            =&& 0,\qquad\quad {\rm for}\quad |t|> l,
\EEA
where $l$ is some parameter that is {\it finite} in the thermodynamical
limit $L\to\infty$. Therefore, the noise given by Eq.~(\ref{srika}) is
obviously finite-range correlated.
Eq.~(\ref{drastamat}) now reads
\BEA
\label{dakar}
\langle \x\rangle=\frac{\sigma}{1-\alpha}\, l^{1-\alpha}\,f^{-2}.
\EEA
If one chooses to take $l\sim \langle \x\rangle$ then $\langle
\x\rangle \sim f^{-2/\alpha}$ as predicted in Ref.~\cite{nelson}.  
However, there is no any a priori reason for this choice, and at any rate
this result refers to the finite-range correlated noise $K_{\rm fr}$.
The real long-range correlated situation, where $l\sim L$, is still not
described by it.

\section{Long-range correlated situation: the frozen noise limit.} 
\label{total}

The present and the next section are devoted to the long-range
correlated situation, where according to Eq.~(\ref{korund}) the
autocorrelation function $K(t)$ of the noise has a power-law behavior
with the single characteristic exponent $1>\alpha>0$.

To start with, let us consider the case with $\alpha\to 0$.  The noise
is now completely frozen: $\eta(s)$ in (\ref{d1}) does not depend on
$s$.  This situation is less physical as compared to that with
$\alpha>0$.  However, it is exactly solvable, and one can hope it
catches at least some features of the realistic situation where
$\alpha$ is larger than zero, but certainly smaller than one. This
intuitive expectation will be confirmed later on.

The problem with $\alpha=0$ is easily solved from (\ref{d1}). Moreover,
the exact solution can be obtained for an arbitrary value of $L$: 
\BEA
\label{karma}
\frac{\x}{L}=g[\,\beta L(f+\eta)\,],
\EEA
\BEA
\label{karma0}
g[x]\equiv \frac{1}{x}-\frac{1}{e^{x}-1}.
\EEA 
It is seen that in the thermodynamical limit $L\to\infty$, $g[\,\beta
L(f+\eta)\,]$ behaves as roughly the step-function, 
$g[\,\beta L(f+\eta)\,]\simeq
\theta(-\eta-f)$: for any single realization of the noise there is
a sharp phase transition with a jump at the realization-dependent
point $f=-\eta$. Exactly at this point $f=-\eta$ one has $g(0)=1/2$
and $\langle \x\rangle=L/2$.

Let us now study the behavior of $\langle \x\rangle$. 
Since the noise is completely frozen, the calculation of 
$\langle \x\rangle$ reduces to the averaging over a Gaussian 
variable with dispersion $\sigma$. We have
\BEA
\label{karma1}
\frac{\langle \x\rangle}{L}=\left\langle\, g[\,\beta L(f+\eta)\,]
\,\right\rangle\equiv
\int \frac{\d\eta}{\sqrt{2\pi\sigma}}\,
\exp\left[-\frac{\eta^2}{2\sigma}\right]\,g[\,\beta L(f+\eta)\,]=
\int\frac{\d \xi}{\sqrt{2\pi\sigma\beta^2}}\,
\exp\left[-\frac{(\xi-\beta f)^2}{2\sigma\beta^2}\right]g[\, L\xi\,],
\EEA
where we changed the integration variable as $\xi=\beta(\eta+f)$. 
In the thermodynamical limit $L\to\infty$, we shall obtain for 
$\langle \x\rangle/L$ the main term of order ${\cal O}(L^0)$, and the
first correction to it which will appear to be of order ${\cal O}(1/L)$.
To this end, let us divide the integration in the RHS of
(\ref{karma1}) into three pieces:
\BEA
\label{mi}
\int =\int_{-\infty}^{-2/L}+\int_{-2/L}^{2/L}+\int_{2/L}^\infty.
\EEA
For each piece we shall use the following aproximate expressions
obtained from (\ref{karma1})
\BEA
\label{korund1}
g[\, L\xi\,]&&=\frac{1}{L\,\xi}, \qquad {\rm for}\qquad
L\,\xi\gtrsim 2,\\
\label{korund2}
g[\, L\xi\,]&&=\frac{1}{2},\qquad\qquad {\rm for}\qquad
-2\lesssim  L\,\xi\lesssim 2,\\
\label{korund3}
g[\, L\xi\,]&&=1+\frac{1}{L\,\xi}, \quad\qquad {\rm for}\qquad
L\,\xi\lesssim -2.
\EEA
To obtain (\ref{korund1}) and (\ref{korund3}) we neglected terms of
order ${\cal O}(e^{-\beta L|f|})$ that is certainly legitimate in the
thermodynamic limit. For (\ref{korund2}) which corresponds to the
second integration piece in (\ref{mi}), we have taken the value of
$g[\, L\xi\,]$ at $\xi=0$. The boundary points of $L\xi$ were chosen
such as to ensure a continuous matching. However, neither the precise value of
$g[\, L\xi\,]$ within the second piece of integration in (\ref{mi}), 
nor the precise values of the points separating this piece from the
remaining ones are important,
since as we show below the contribution coming from this 
second piece, as well as the contributions from the boundary points of the
two other integration pieces, produce factors of order ${\cal
O}(1/L^2)$ at best.

Combining (\ref{korund1}--\ref{korund3}) with (\ref{karma1}) one gets
\BEA
\frac{\langle \x\rangle}{L}&&=
\int_{-\infty}^{-2/L}\frac{\d \xi}{\sqrt{2\pi\sigma\beta^2}}\,
\exp\left[-\frac{(\xi-\beta f)^2}{2\sigma\beta^2}\right]
\left(1+\frac{1}{L\xi}\right)+\frac{1}{2}
\int_{-2/L}^{2/L}\frac{\d \xi}{\sqrt{2\pi\sigma\beta^2}}\,
\exp\left[-\frac{(\xi-\beta f)^2}{2\sigma\beta^2}\right]\nonumber\\
\label{taras}
&&+\int_{2/L}^\infty\frac{\d \xi}{\sqrt{2\pi\sigma\beta^2}}\,
\exp\left[-\frac{(\xi-\beta f)^2}{2\sigma\beta^2}\right]
\,\frac{1}{L\xi}\\
\label{taras10}
&&=\int_{-\infty}^{0}\frac{\d \xi}{\sqrt{2\pi\sigma\beta^2}}\,
\exp\left[-\frac{(\xi-\beta f)^2}{2\sigma\beta^2}\right]+
\frac{1}{L}\int_{0}^{\infty}\frac{\d \xi}{\xi\sqrt{2\pi\sigma\beta^2}}\left(
\exp\left[-\frac{(\xi-\beta f)^2}{2\sigma\beta^2}\right]
-\exp\left[-\frac{(\xi+\beta f)^2}{2\sigma\beta^2}\right]
\right)\\
\label{taras11}
&&-\frac{1}{L}\int_{0}^{2/L}\frac{\d \xi}{\xi\sqrt{2\pi\sigma\beta^2}}\left(
\exp\left[-\frac{(\xi-\beta f)^2}{2\sigma\beta^2}\right]
-\exp\left[-\frac{(\xi+\beta f)^2}{2\sigma\beta^2}\right]
\right)\\
&&+\half \int_{0}^{2/L}\frac{\d \xi}{\sqrt{2\pi\sigma\beta^2}}\left(
\exp\left[-\frac{(\xi-\beta f)^2}{2\sigma\beta^2}\right]
-\exp\left[-\frac{(\xi+\beta f)^2}{2\sigma\beta^2}\right]
\right).
\label{taras12}
\EEA
One notes that both (\ref{taras11}) and 
(\ref{taras12}) are of order ${\cal O}(1/L^2)$. This can be verified by
directly expanding integrals in (\ref{taras11}) and 
(\ref{taras12}) for small $2/L$. Skipping these terms, one gets
\BEA
\frac{\langle \x\rangle}{L}&&=
\int_{\beta f}^{\infty}\frac{\d \xi}{\sqrt{2\pi\sigma\beta^2}}\,
\exp\left[-\frac{\xi^2}{2\sigma\beta^2}\right]
+\frac{1}{L}
\int_0^{\infty}\frac{\d \xi}{\xi\,\sqrt{2\pi\sigma\beta^2}}\,
\left(\exp\left[-\frac{(\xi-\beta f)^2}{2\sigma\beta^2}\right]
-\exp\left[-\frac{(\xi+\beta f)^2}{2\sigma\beta^2}\right]
\right)\nonumber\\
&&=\int_{\beta f}^{\infty}\frac{\d \xi}{\sqrt{2\pi\sigma\beta^2}}\,
\exp\left[-\frac{\xi^2}{2\sigma\beta^2}\right]
+\frac{1}{\sigma\beta^2 L}
\exp\left[-\frac{f^2}{2\sigma}\right]\int_0^{\beta f}\d \xi\,
\exp\left[\frac{\xi^2}{2\sigma\beta^2}\right],
\label{bulba}
\EEA
When obtaining the last term in the RHS of (\ref{bulba}), we used
a tabulated identity for the error function.

For $f$ not very large as compared to $\sqrt{\sigma}$, the first term
in the LHS of (\ref{bulba}) is dominating: ${\langle \x\rangle}/{L}$
is or order one-half. In particular, it is exactly equal to one-half
for $f=0$. The dependence of $\langle\x\rangle$ on $f$ becomes thus
very weak for $f\to 0$.  The second, subdominant term become
non-negligible for $f\gg \sqrt{\sigma}$, where using asymptotic
identities (see Appendix \ref{aux}):
\BEA
\int_{a}^\infty\frac{\d\xi}{\sqrt{2\pi}}\,e^{-\xi^2/2}=
\frac{e^{-a^2/2}}{a\sqrt{2\pi}}\,\left(1-\frac{1}{a^2}+\dots\right),
\qquad a\gg 1,
\EEA
\BEA
\int_0^{a}{\d\xi}\,e^{\xi^2/2}=
\int_0^{a}{\d\xi}\,e^{\xi^2/2}=
\frac{e^{a^2/2}}{a}\left(1+\frac{1}{a^2}\right)+
\dots,\qquad a\gg 1,
\EEA
one gets from (\ref{bulba}) noting $a=f/\sqrt{\sigma}$:
\BEA
{\langle \x\rangle}=\frac{1}{\beta f}\left(1
+\frac{\sigma}{f^2}
+\frac{\sqrt{\sigma}}{\sqrt{2\pi}}\,L
\beta\,\exp\left[-\frac{f^2}{2\sigma}\right]
\right).
\label{kaban}
\EEA
Note that for $f\gg \sqrt{\sigma}$, $\langle\x\rangle$ has | within
the leading order | the same $1/f$-dependence as it will be in the
completely homogeneous situation without noise ($\sigma=0$). In the
considered regime, the noise only renormalizes this behavior modifying
the subdominant terms.

\begin{figure}[bhb]
\includegraphics[width=10cm]{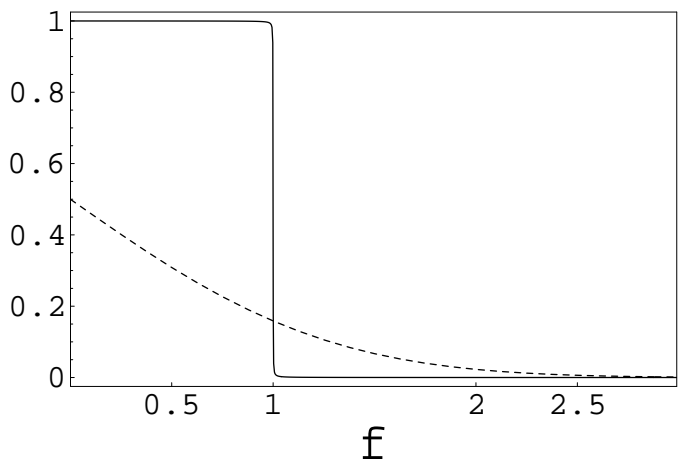}
\caption{
$\x(f)/L$ for a particular realization (solid curve) 
and $\langle\, \x(f)\,\rangle/N$ (dotted curve) versus $f$;
$T=\sigma=1$, and $L=10^{4}$. }
\label{f04}
\end{figure}

As compared to $\x/L$ which has a jump at a realization dependent
point $f=\eta$, ${\langle \x\rangle}/{L}$ is seen to behave smoothly.
It displays a crossover between small $\langle \x\rangle/L$ for a
large $f$ and $\langle \x\rangle=L/2$ for $f=0$: the sharp transition
disappears; see Fig.~\ref{f04}. This indicates that the situation for
the totally-correlated noise is essentially non-self-averaging: in the
thermodynamical limit the averaged order parameter $\langle \x\rangle$
does not reproduce the behavior of $\x$ for a typical realization.
Recall that for disordered systems all observables like free energy,
order parameters, correlation functions, etc., depend on the
realization of the disorder, i.e. they are random quantities. It is of
the immediate interest to know their most probable (typical) values,
since they will be met in experiments. If for a given quantity its
typical value in the thermodynamic limit coincides with its average,
one speaks on self-averaging; see e.g. \cite{brout,de}. In practice
this means that it is sufficient to study averages as they are
representative in the single sample measurements. It is known on the
general ground that in the proper thermodynamic limit, that is when
the linear size $L$ of the system is much larger than any other
characteristic length and provided the distribution of the disorder is
finite-range correlated, quantities that scale with the volume of the
studied random system | these are extensive quantities such as free
energy, order parameter, but not the statistical sum | are expected to
display self-averaging \cite{brout,de}. This result is based on the
law of large numbers.  However, this need not be true if the
distribution of the disorder is long-range correlated, since now the
correlation length of the disorder has the same order of magnitude as
the linear size, and the arguments based on the law of large numbers
do not apply. The above situation is just of this sort.

\subsection{Dispersion as a measure of non-self-averaging.}

It is desirable to have more quantitative indications of the above
indicated non-self-averaging effect. To characterize fluctuations 
of $\x$ from one realization to another, it is natural to employ the
corresponding dispersion $\langle \x^2\rangle-\langle \x\rangle^2$
which tells us how the quantity $\langle \x\rangle$ fluctuates from
one realization to another. Then the statement of self-averaging will
read:
\BEA
\label{said}
\frac{\langle \x^2\rangle-\langle \x\rangle^2}{\langle \x\rangle^2}
\to 0, \qquad {\rm for}\qquad L\to \infty.
\EEA
In contrast, if ${(\langle \x^2\rangle-\langle
\x\rangle^2)}/{\langle \x\rangle^2}$ remains finite for $L\to \infty$,
we have non-self-averaging.

The quantity $\langle \x^2\rangle$ can be calculated in the same way
as in Eqs.~(\ref{taras}, \ref{bulba}). We shall bring the result
only for $f$ not very large as compared to $\sqrt{\sigma}$, that is,
when $\langle \x^2\rangle \propto L^2$:
\BEA
\frac{\langle \x^2\rangle}{L^2}=
\int_{\beta f}^{\infty}\frac{\d \xi}{\sqrt{2\pi\sigma\beta^2}}\,
\exp\left[-\frac{\xi^2}{2\sigma\beta^2}\right]
+\frac{2}{L}\,\ln \left(\frac{L}{2}\right)\,~
\frac{\exp\left[-\frac{f^2}{2\sigma}\right]}{\sqrt{
2\pi \beta^2\sigma}}+{\cal O}\left(\frac{1}{L}\right).
\label{mulba}
\EEA
Substituting this into (\ref{said}), we see that 
${(\langle \x^2\rangle-\langle \x\rangle^2)}/{\langle \x\rangle^2}$
remains finite in the thermodynamical limit: 
\BEA
\frac{\langle \x^2\rangle-\langle \x\rangle^2}{\langle \x\rangle^2}
=\frac{\int_{\beta f}^{\infty}\frac{\d \xi}{\sqrt{2\pi\sigma\beta^2}}\,
\exp\left[-\frac{\xi^2}{2\sigma\beta^2}\right]}
{\left(\int_{\beta f}^{\infty}\frac{\d \xi}{\sqrt{2\pi\sigma\beta^2}}\,
\exp\left[-\frac{\xi^2}{2\sigma\beta^2}\right]
\right)^2}
\,-\,1.
\EEA
In particular, for $f\to 0$
\BEA
\label{kaput}
\frac{\langle \x^2\rangle-\langle \x\rangle^2}{\langle \x\rangle^2}\to
1,
\EEA
indicating essential non-self-averaging.

In closing this section, let us repeat that the character of the
thermodynamical for the considered case $\alpha=0$ is different from
that of the finite-range correlation situation, where | for $L\to
\infty$ | the behavior of $\langle\x\rangle$ become independent on $L$
at least in the physical range of other parameters (e.g., $1>f>0$,
$T>0$, etc).  For the $\alpha=0$ case, as seen from Eqs.~(\ref{bulba},
\ref{kaban}), there is an explicit dependence on $L$ in the whole
range of physical range of the involved parameters.  According to
(\ref{kaban}), if $L$ is kept large but finite, then this dependence
is very weak for external forces far form their critical value $f=0$,
that is, for $f\gg\sqrt{\sigma}$. There are no reasons for taking this
explicit dependence on $L$ as something unphysical. In contrast, the
actual size of physically relevant examples of DNA is never more that
$L\sim 10^4-10^6$; see Ref.~\cite{bio}. This is certainly much smaller
than the number $10^{23}$ which in the standard statistical physics is
taken as the typical size. Therefore, it is rather natural to study the
physics of unzipping for a large but fixed $L$.

For the considered frozen situation, we could solve the
problem analytically for a given realization. However, for $\alpha>0$
this is not possible, and one has to rely on numerical methods. This
is what we intend to do in the next section.

\begin{figure}[b]
\includegraphics[width=10cm]{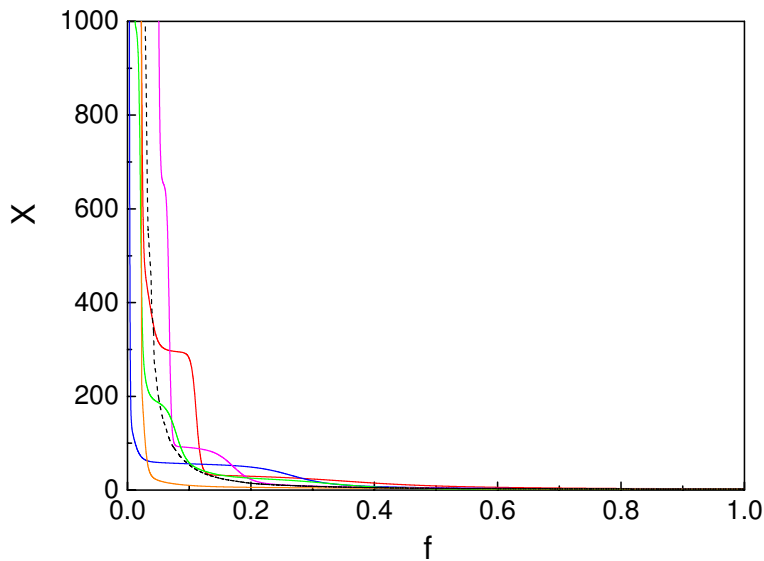}
\caption{Solid curves: $\x(f)$ for several realizations of the white
uncorrelated noise.  Dotted curve: $\langle \x(f)\rangle$ obtained by
averaging over $10^3$ realizations.  $T=D=1$, $L=5\times 10^{4}$. 
}
\label{fig_1a}
\end{figure}

\begin{figure}[t]
\includegraphics[width=10cm]{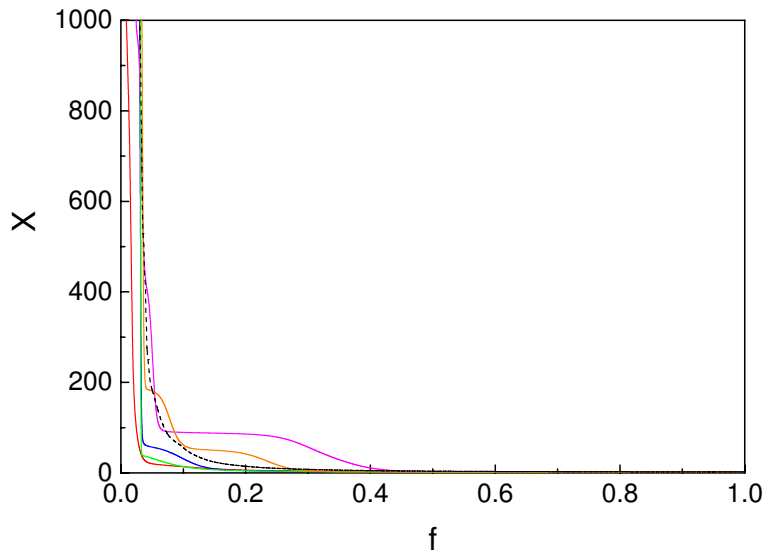}
\caption{ Solid curves: $\x(f)$ for several realizations of the
dichotomic uncorrelated noise ($\eta_i=\pm 1$ with equal probability).
Dotted curve: $\langle \x(f)\rangle$ obtained by averaging over $10^3$
realizations.  $T=1$, $L=5\times 10^{4}$. 
}
\label{fig_1b}
\end{figure}


\begin{figure}[t]
\includegraphics[width=12cm]{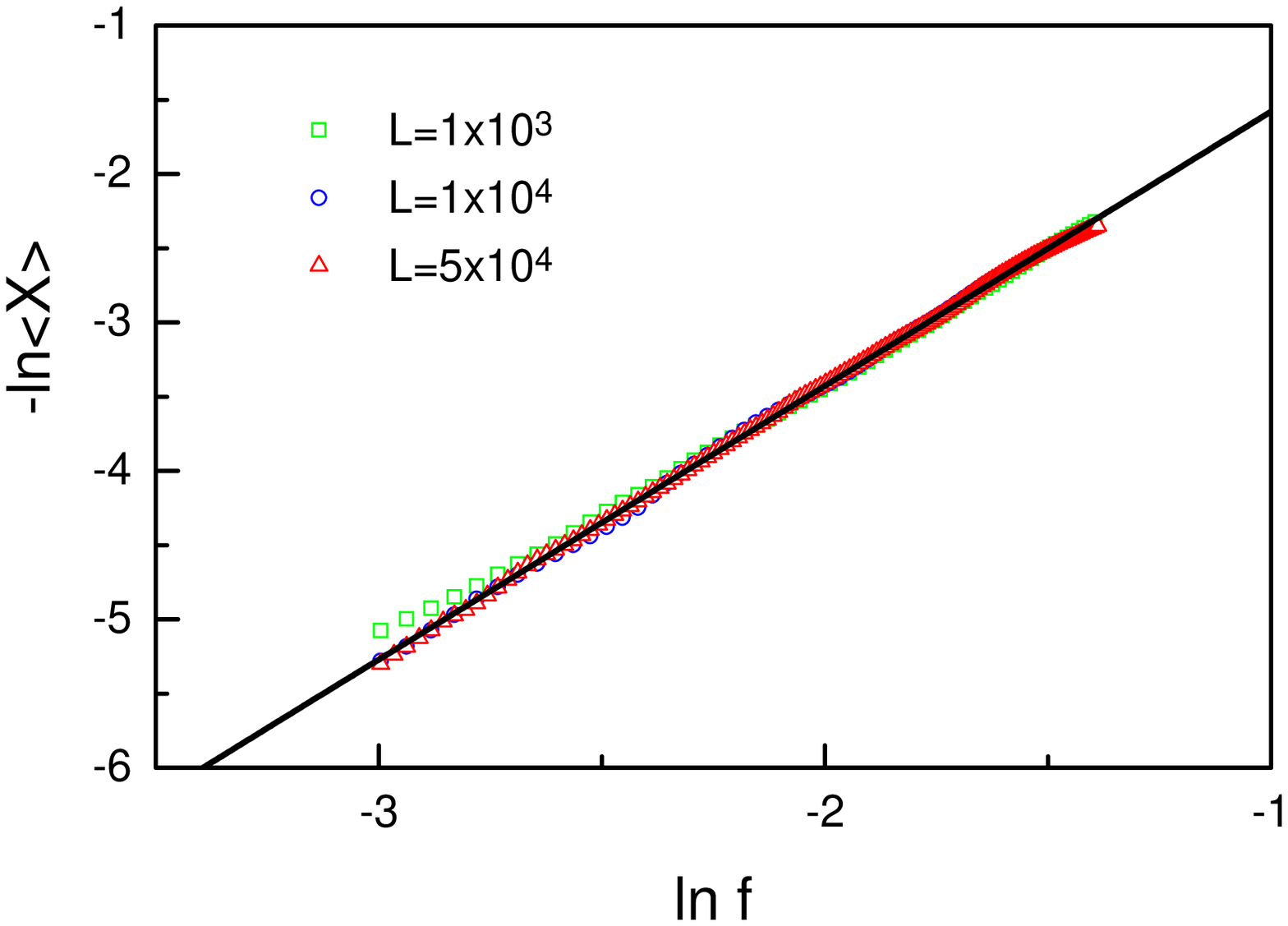}
\caption{
The dependence of $-\ln \langle\x(f)\rangle$ on $\ln f$ for various
values of $L$ and for $T=D=1$.  The quantity $\langle\x(f)\rangle$ was
obtained by direct averaging over $10^3$ realizations. Solid line
idicates linear fitting $-\ln \langle\x(f)\rangle=A+1.84179\,\ln f$
for $L=5\times 10^4$,
where $A$ is a constant.  The emergence of the power-law (\ref{klen})
is thus displayed explicitly. 
}
\label{fig_2-1b}
\end{figure}

\section{Numerical results.}
\label{typical}

As we have seen in the previous section, there are reasons to expect
that for the long-range correlated situation, especially for
sufficiently small index $\alpha$, the typical | that is, frequently
met among many independent realizations of the noise | behavior of
$\x(f)$ in the thermodynamic limit is not described adequately by the
average quantity $\langle \x\rangle$ (non-self-averaging).  We note in
this context that the correlator $(\langle \x^2\rangle-\langle
\x\rangle^2)/{\langle \x\rangle^2}$ studied in section \ref{total} can
indicate on non-self-averaging, but by itself does not provide any
direct information on typical realizations. It is perhaps needless to
stress that once we expect the effect of non-self-averaging, the
attention should be shifted towards typical realizations, since they
do have a direct physical meaning for single-molecule experiments.


In the present section we study numerically the behavior of the number
of broken base-pairs $\x$ as a function of $f$ both for the long-range
correlated situation and for the uncorrelated noise.  For the discrete
version of the model the partition function reads:
\BEA
\label{zanzibar}
Z=\sum_{k=1}^L\exp[-\beta(fk+\sum_{i=1}^k\eta_i)], 
\EEA 
where for the long-range correlated situation $\eta_i$ are Gaussian
random variables with the autocorrelation function given by
(\ref{korund}). Note that for the purposes of numerical computations
the behavior of $K(t)$ in (\ref{korund}) was regularized at short
distances so as to avoid superfluous short-range singularities; see
Appendix \ref{num} for details.  The generation of $\eta_i$,
$i=1,...,L$ is described in Appendix \ref{num} following to optimized
recipes proposed in \cite{sancho}.  For numerical computations we have
chosen $T=1$ and $L=10^4$ or $L=5\times 10^4$.

As $L$ is now explicitly finite, one should be careful with
the selection of the thermodynamical domain, since due to the very
statement of the problem the limit $L\to\infty$ is taken before $f\to
0$. As a plausible estimate of this domain, one can use a condition
$f\sqrt{L}\gg 1$. We confirmed it in several ways, reproducing 
predictions which were made in the thermodynamica limit $L\to\infty$.

\subsection{Uncorrelated noise.}
\label{un}




\begin{figure}[t]
\includegraphics[width=10cm]{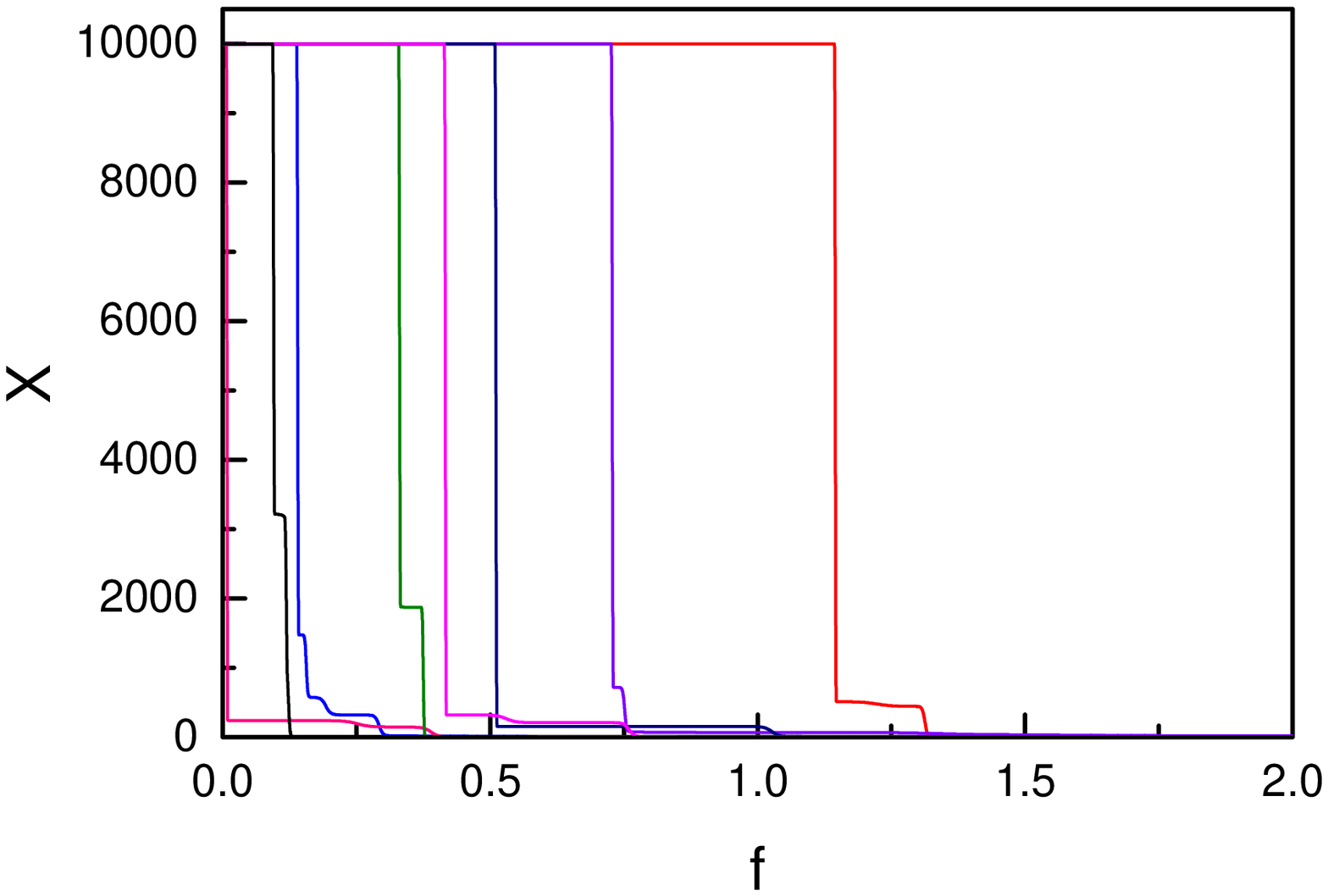}
\caption{
Realizations of $\x(f)$ from the first class of typicality.
$T=\sigma=2\alpha=1$, $L=10^4$.
}
\label{fig_3-11a}
\end{figure}

\begin{figure}[b]
\includegraphics[width=10cm]{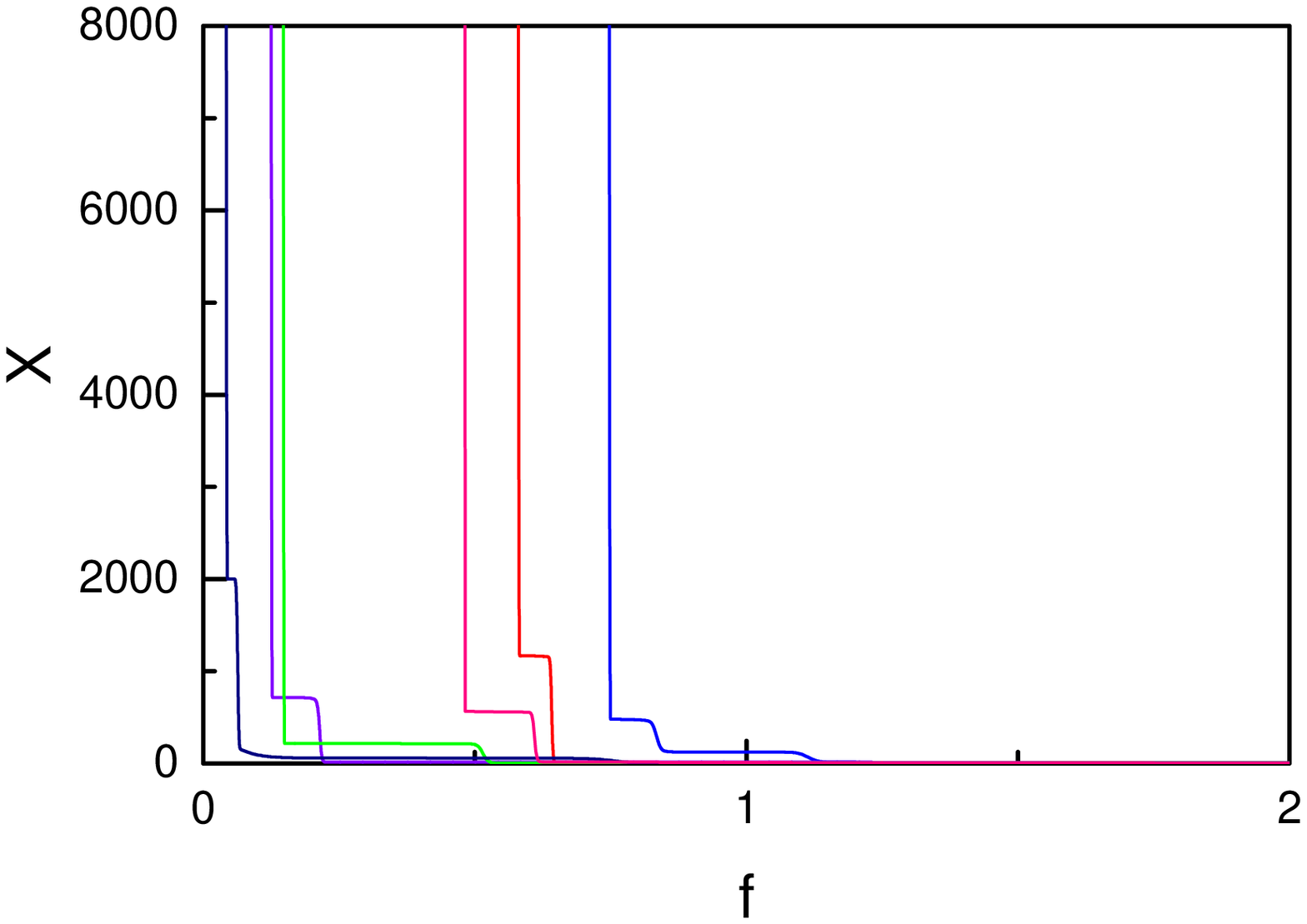}
\caption{Realizations of $\x(f)$ from the first class of typicality.
$T=\sigma=2\alpha=1$, $L=5\times 10^4$.
}
\label{fig_3-2a}
\end{figure}

\begin{figure}[t]
\includegraphics[width=10cm]{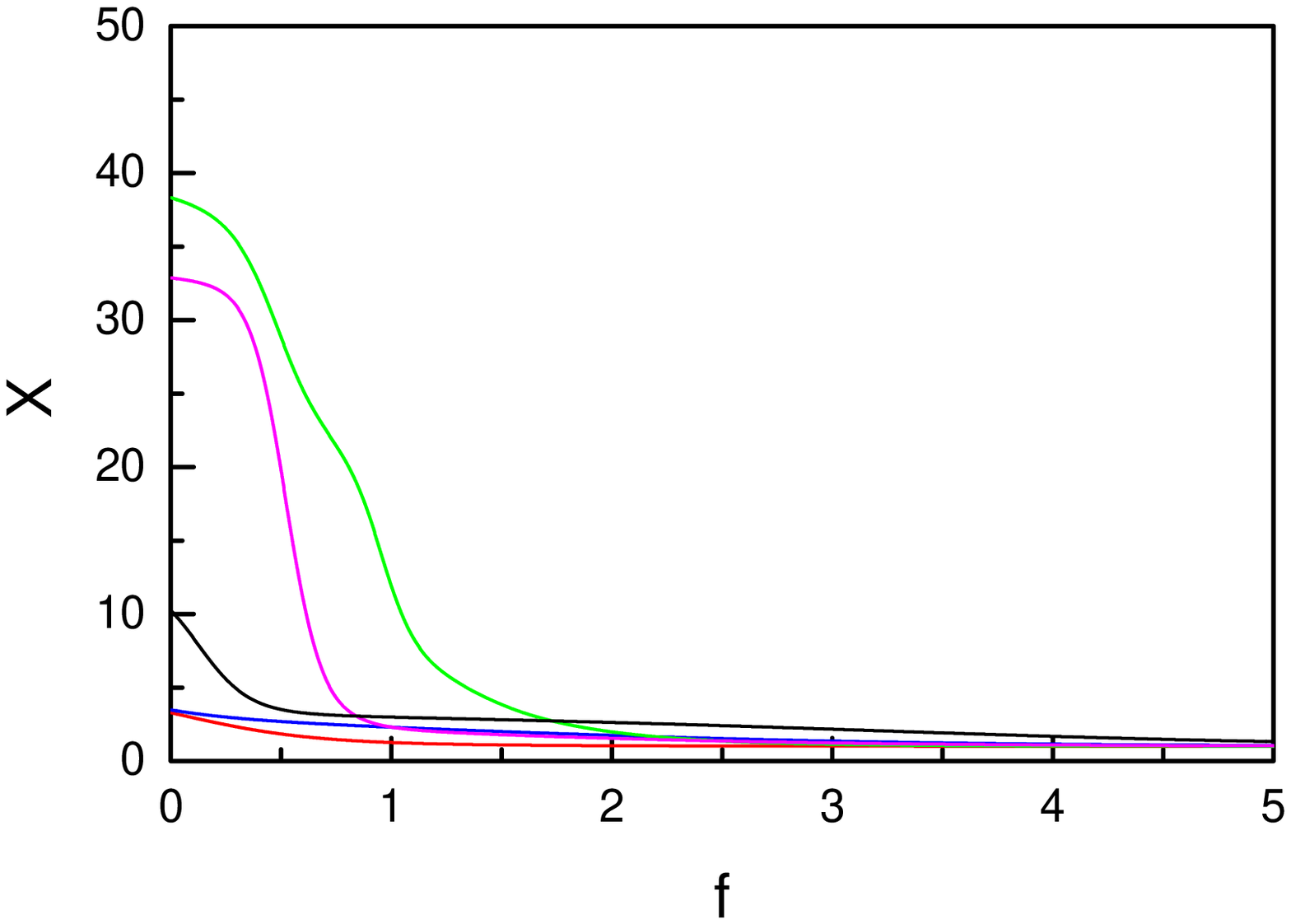}
\caption{Realizations of $\x(f)$ from the second class of typicality.
$T=\sigma=2\alpha=1$, $L=10^4$.
}
\label{fig_3-1b}
\end{figure}

\begin{figure}[b]
\includegraphics[width=10cm]{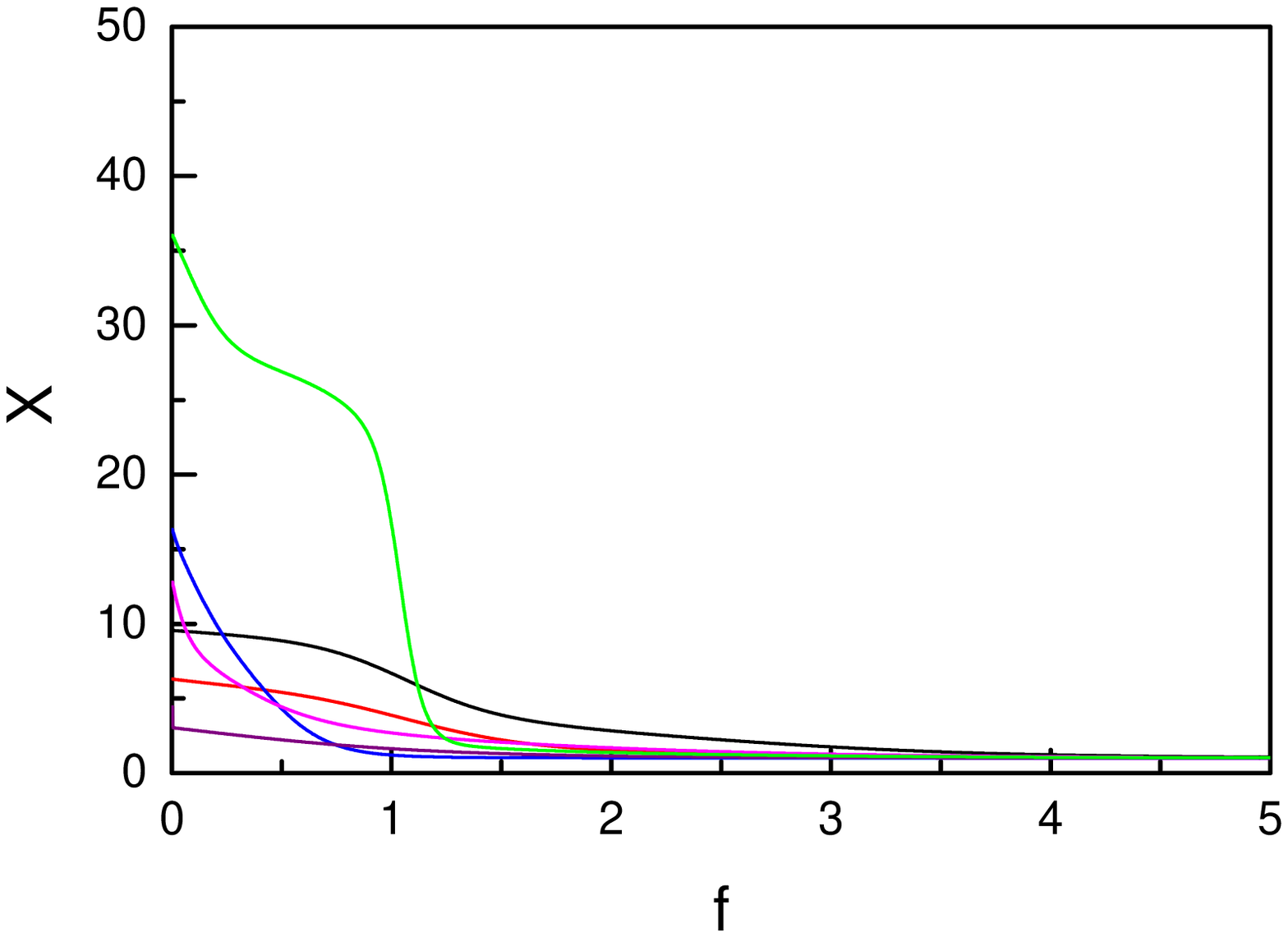}
\caption{Realizations of $\x(f)$ from the second class of typicality.
$T=\sigma=2\alpha=1$, $L=5\times 10^4$.
}
\label{fig_3-2b}
\end{figure}

Let us start with the uncorrelated noise case, where $\eta_i$'s are
independent Gaussian variables with zero average,
$\langle\eta_i\rangle=0$, and variance $\langle\eta^2_i\rangle=D=1$
(white noise), and where $X$ is given by (\ref{brams},
\ref{zanzibar}).  For comparison we also studied a case, where $\eta_i$
are independent random variables assuming values $\eta_i=\pm 1$ with
equal probability (dichotomic noise).

The results are illustrated by Figs.~(\ref{fig_1a},
\ref{fig_1b}), where we display $\langle\x\rangle$ and $\x$ for
several typical realizations. It is seen that $\langle\x\rangle$ and
$\x$ do not coincide exactly, as it is in general expected due to the
finite magnitude of $L$ if not by any other reason. However, in the
considered thermodynamical domain of $f$ the behavior of various
typical realizations qualitatively resembles each other, and,
therefore, resembles that of $\langle \x(f)\rangle$.  In particular,
for all typical realizations $\x(f)$ grows for $f\to 0$.  In that
sense $f=0$ is a special point for both typical $\x$ and
$\langle\x\rangle$. It should be mentioned that for $f\le 0.05$ we
have seen realizations containing relatively sudden jumps at
realization-dependent values of $f$. This differs from the behavior of
$\langle\x\rangle$ and is in agreement with results of Ref.~\cite{nelson}.
However, such small values of $f$ are not in the thermodynamical
domain. Acknowledging reservations connected with the numerical
character of our study, we, nevertheless, conclude that the
uncorrelated-noise situation is self-averaging at least for not very small,
$f\sqrt{L}\gg 1$, values of $f$.

It is seen from Fig.~(\ref{fig_1b}) that the white and dichotomic
noise produce very similar results.  This is to be expected for the
considered large values of $L$ (law of large numbers).  
Fig.~(\ref{fig_2-1b}) shows that the power law $\langle\x\rangle \propto
f^{-2}$ for the white-noise case is recovered by direct averaging over
various realizations. Indeed, it is seen from this figure that one recovers
\BEA
\label{klen}
\langle\x\rangle
\propto f^{-1.84},
\EEA
after averaging over $10^3$ realizations in the domain
$0.05<f<0.25$. This result is stable upon increasing the number of
realizations, e.g. from $10^3$ to $2\times 10^3$.

\subsection{Long-range correlated noise}

\begin{figure}[t]
\includegraphics[width=10cm]{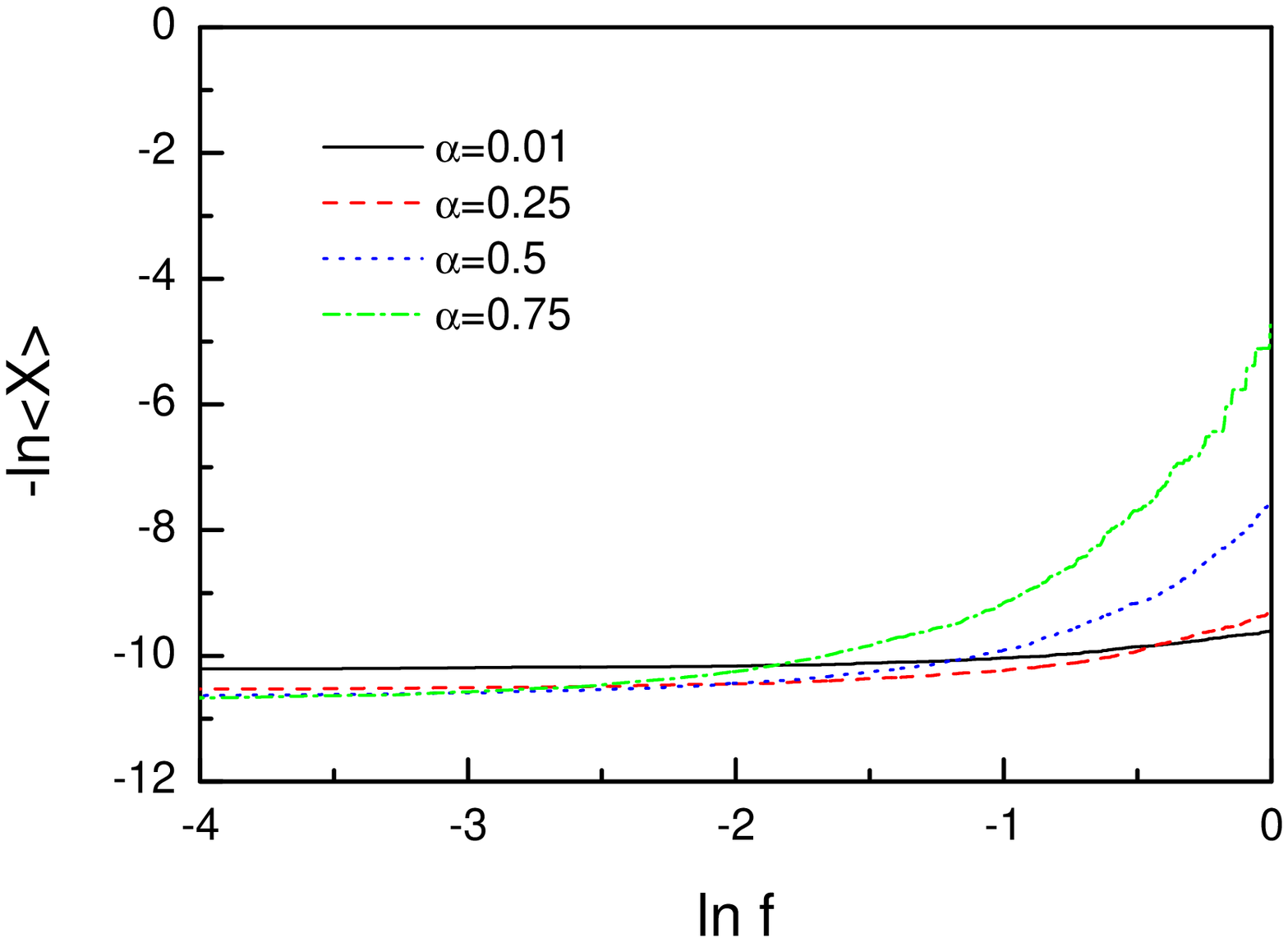}
\caption{$-\ln\langle \x\rangle$ versus $\ln f$ for the long-range
correlated noise with various $\alpha$'s and $L=5\times 10^4$,
$T=\sigma=1$.
The quantity $\langle\x(f)\rangle$ was obtained by direct 
averaging over $10^3$ realizations.
}
\label{fig_6-2a}
\end{figure}

\begin{figure}[b]
\includegraphics[width=10cm]{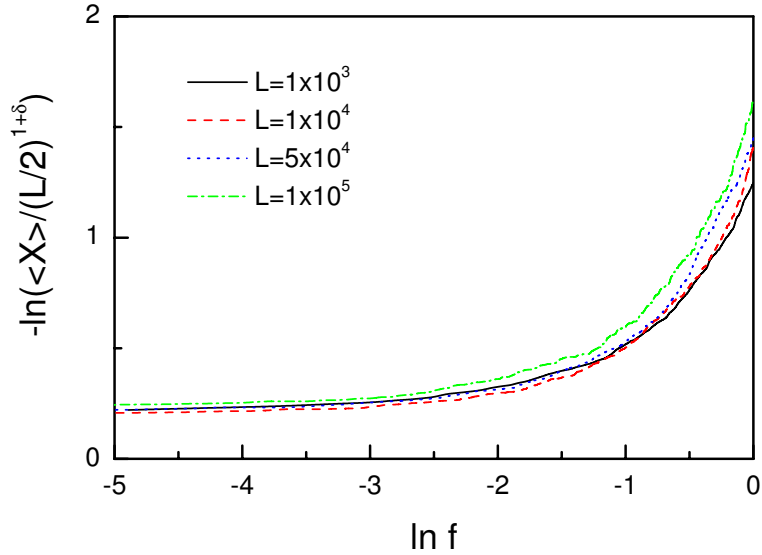}
\caption{$-\ln\left(\langle\x\rangle\,(L/2)^{-1-\delta(\alpha)} \right)$
versus $\ln f$ for various $L$'s and $T=\sigma=1$, $\alpha=0.25$,
$\delta=0.0625$. The quantity 
$\langle\x(f)\rangle$ was obtained by direct 
averaging over $10^3$ realizations.
}
\label{fig_7-2b}
\end{figure}

\begin{figure}[t]
\includegraphics[width=10cm]{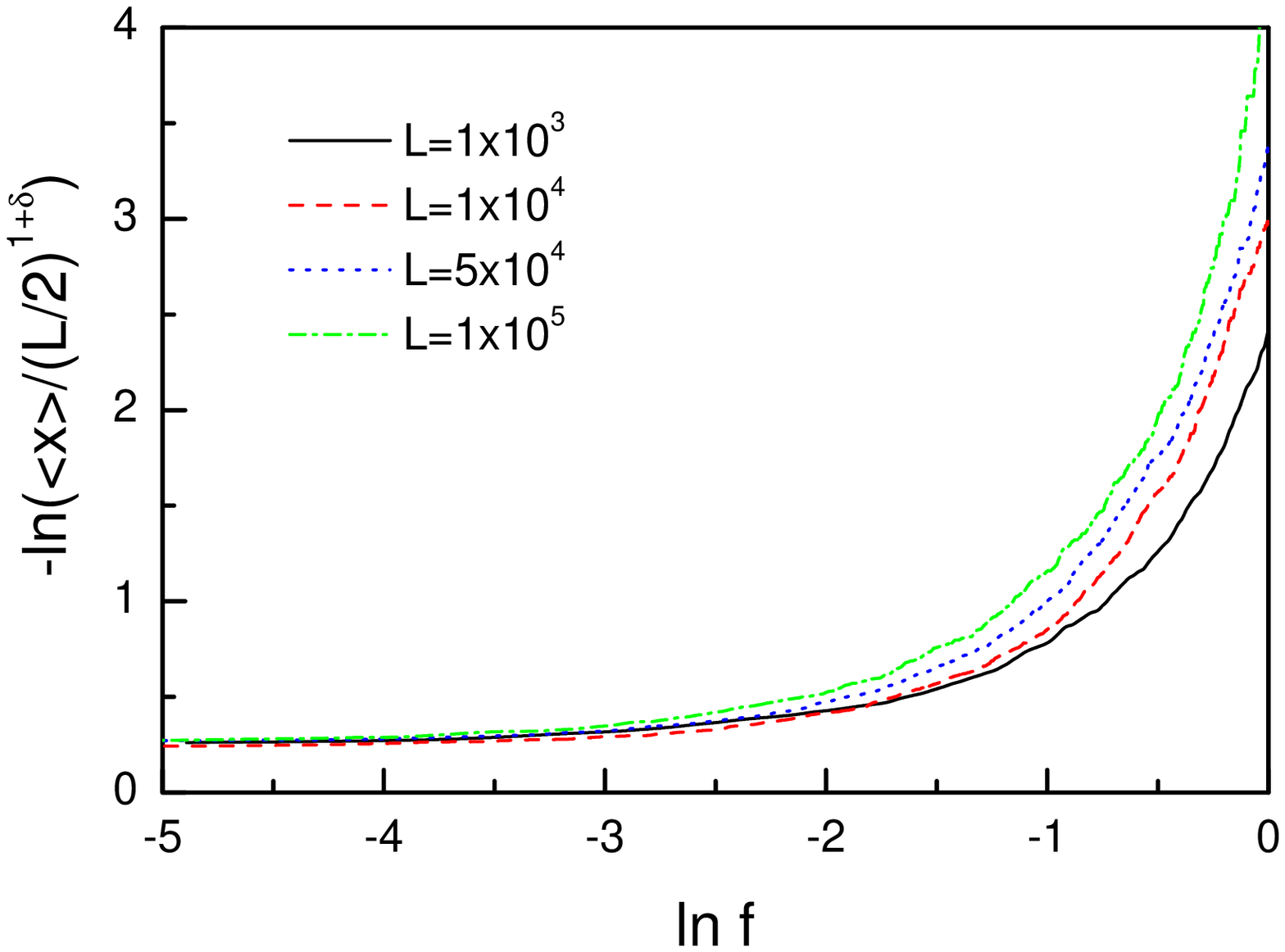}
\caption{The same as in Fig.~(\ref{fig_7-2b}) but with 
$\alpha=0.5$ and $\delta=0.075$.
}
\label{fig_7-3b}
\end{figure}

\begin{figure}[b]
\includegraphics[width=10cm]{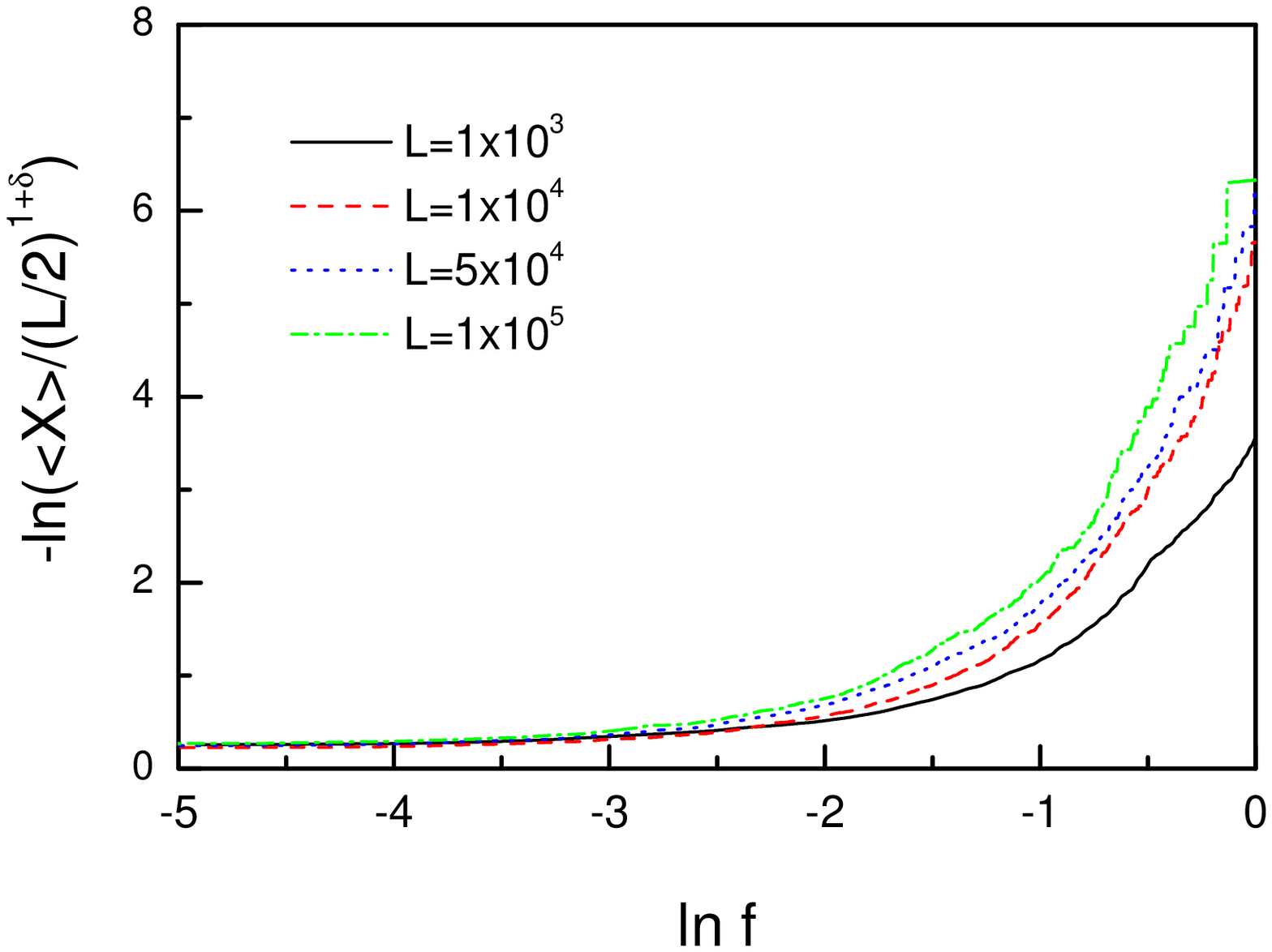}
\caption{The same as in Fig.~(\ref{fig_7-2b}) but with 
$\alpha=0.75$ and $\delta=0.08$.
}
\label{fig_7-4b}
\end{figure}

\begin{figure}[t]
\includegraphics[width=10cm]{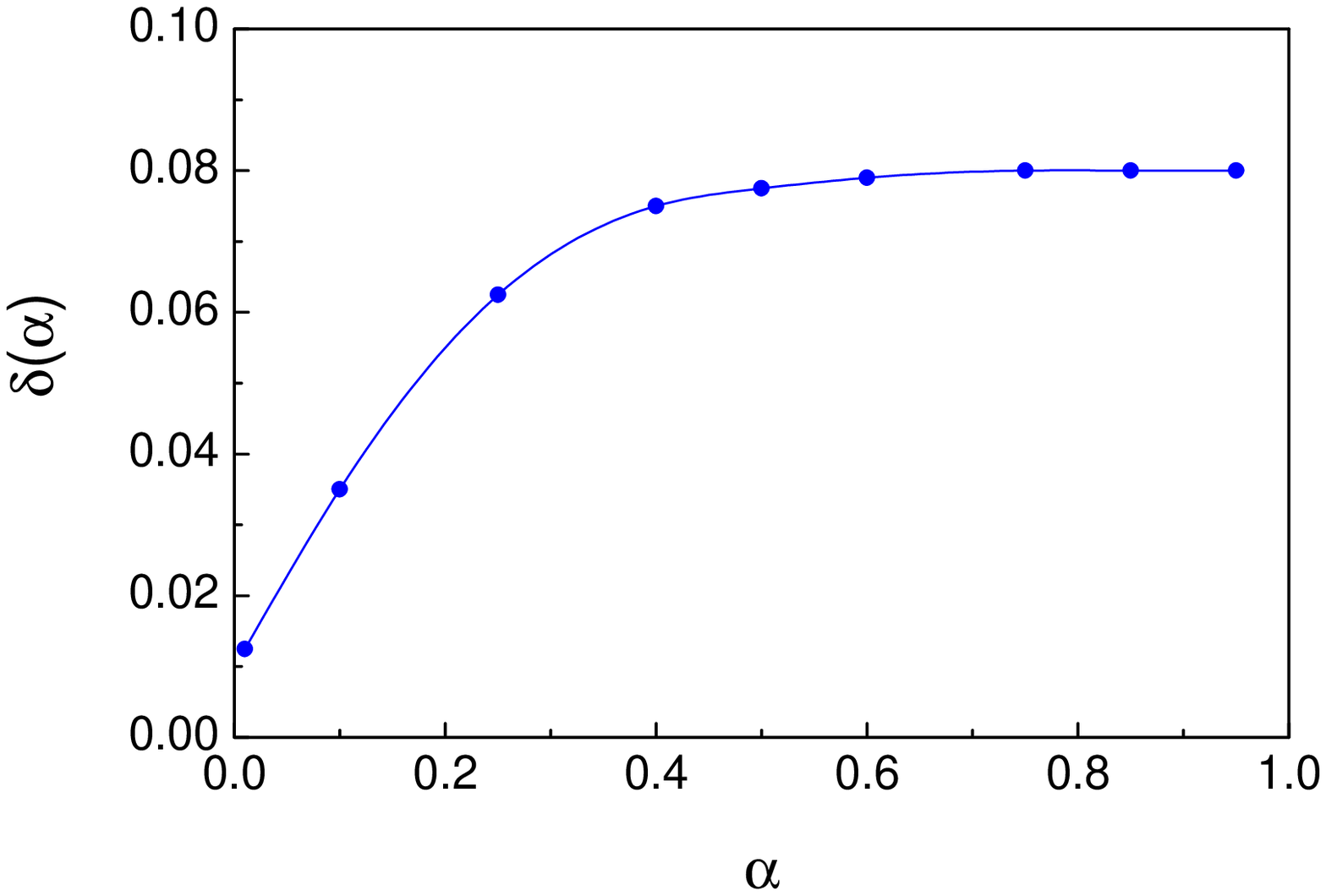}
\caption{$\delta(\alpha)$ defined by Eq.~(\ref{tiratsu}) versus
$\alpha$.
}
\label{fig_8}
\end{figure}

\subsubsection{Typical realizations.}

The situation for the long-range correlated noise for $\alpha=0.5$,
$T=\sigma=1$ is illustrated by Figs.~(\ref{fig_3-11a},
\ref{fig_3-1b}, \ref{fig_3-2a}, \ref{fig_3-2b}). The first point to
note is that now there are typical realizations with radically
different properties. The first type of realizations is presented by
Figs.~(\ref{fig_3-11a}, \ref{fig_3-2a}): $\x(f)$
increases by several sudden jumps followed by flat regions. It is seen
that $\x|_{f=0}$ is either equal to its maximal possible value $L$ or
is close to it. Points where $\x(f)$ has jumps vary from one
realization to another. However, the overall number of jumps when
varying $f$ between zero and one is typically two or three.

In contrast to this, Figs.~(\ref{fig_3-1b}, \ref{fig_3-2b})
presents a strictly different situation: It is characterized by very
smooth behavior of $\x(f)$ for $f\ge 0$. In particular, $\x|_{f=0}$ is
much smaller than $L$ (typically by few orders of magnitude).  
$\x(f)$ is still a monotonic function of $f$, but the point $f=0$ |
where the energy supplied by the external unzipping force is equal to
the {\it average} binding energy of a base-pair | is by no means
special.

To estimate the frequency by which each scenario is met among all
possible realizations, we have taken the following criteria for
deciding whether a given realization belongs to one of the above
classes: for $L=5\times 10^4$ we prescribe the given realization to
the first class if $\x(f=0)>4.8\times 10^4$, while it is prescribed to
the second class if $\x(f=0)<10^2$. These criteria appeared to be
sufficiently adequate, as they are consistent with the fact of
presence (for the first class) of absence (for the second class) of
sudden jumps for $\x(f)$. 

In this way the frequencies of each class were estimated in a sample
of $10^3$ realizations. It appeared for $L=5\times 10^4$ and
$T=\sigma=2\alpha=1$ that the first scenario is met in $\sim 84\%$ of
all cases ($839$ in $10^3$ realizations), while the second scenario is
present in $\sim 12\%$ of all cases ($118$ in $10^3$ realizations).
These fractions are stable upon increasng the size of the sample on
which the above estimation were carried out.  Interestingly enough,
realizations where $\x(f)$ as a function of $f$ fall into neither of
the above two classes amount only to $\sim 4\%$ of all possible cases.

It is relevant to note that the fractions of the two classes
show tendency to move towards each other upon decreasing the size of
the system. For instance, the fractions of the first and the second
class amount to $18\%$ and $76\%$, respectively, for $L=10^4$
($T=\sigma=2\alpha=1$). These fractions were estimated by criteria
$\x(f=0)>0.8\times 10^4$ and $\x(f=0)< 10^2$, respectively.  Recall in
this context that the chosen values for $L$ are sensible, since the
typical DNA samples used in experiment have $L\sim 10^4-10^5$; see, e.g.,
\cite{bio,biofiz} and also section \ref{infer}.

Since there are typical realizations which are so much different from
each other, we conclude that this long-range situation is essentially
non-self-averaging in the whole physical domain $0<f<1$ and, in
particular, in the thermodynamical domain of $f$. This fact
distinguishes between the uncorrelated (white noise) and long-range
correlated situations.  It should be noted that due to the law of
large numbers any non-self-averaging present in the whole domain
$0<f<1$ is certainly impossible for the uncorrelated (or weakly
correlated) noise \cite{brout,de}. For the long-range correlated case
the very law of large numbers does not apply, and the above effect
becomes possible.

Our discussion of the frozen noise presented in section \ref{total}
allows to provide a qualitative explanation for features of the above
two classes of typical realizations. One notes that a sizeable portion
of long-range correlated noise realizations can be seen as several
pieces of the frozen noise with different $\eta$'s put next to each
other.  Now recall from (\ref{karma}) that every sufficiently long
piece of that type has a single first order phase transition with a
jump proportional to its length.

The same reasoning can be applied for the understanding of the
existence of the second class, where $\x(f)$ is a smooth function of
$f$ and $\x(f=0)\ll L$. Here one should note that | within the above
qualitative image of a long-range correlated random sequence | there
are realizations of the noise where all $\eta$'s are positive, and
thus all jumps of $\x(f)$ can occur only for negative $f<0$, that is,
beyond the domain of our interest.

\subsubsection{Inferring phase transitions.}
\label{infer}

Let us finally discuss on whether we can infer phase transitions by
studying the typical realizations. First of all, it is obvious that
once we do not have self-averaging, phase transitions should be
studied on typical scenarios of behavior for $\x$ and not on the
behavior of its average $\langle\x \rangle$. There is another aspect
which is certainly more subtle: phase transitions are typically
defined in the thermodynamical limit and one needs special tools of
finite-size scaling for their identification in results of numerical
computations which are necessarily done on finite systems. The idea of
the finite-size scaling is thus to extrapolate these results to the
thermodynamical limit.  However, there is another, somewhat different
line of thought \cite{gross} which identifies the proper
thermodynamical quantities (such as entropy, free energy, order
parameter, etc) {\it directly} for finite systems, and then searches
in the space of parameters some points having a special character for
this quantities. This approach well-recommended itself for studying
phase transitions in atomic and nuclear physics, and in systems with
long-range interactions (e.g., a gas of self-gravitating particles).
For the present study of DNA there is a related aspect that should be
taken into account: in natural conditions the number of base-pairs is
large, {\it but finite}. Here $L$ is of order of $10^4\sim 10^5$, see
e.g.  \cite{bio,biofiz}, as we have mentioned already.  It is,
therefore, clear that the considered finite size aspect of DNA is
something generic, and not only connected with natural limitations of
numerical methods.

Let us now return to the situation presented in 
Figs.~(\ref{fig_3-1b}, \ref{fig_3-2a}, \ref{fig_3-2b}).  We are going to use
the analogy with the case of the totally correlated noise described in
section \ref{total}. It was seen already that this analogy helped us
to draw useful qualitative conclusions on the numerical data. For the
totaly correlated noise the point of the phase transition is
anambiguously identified with the realization dependent value
$f=-\eta$. At this point the order parameter $\x$ has a jump of order
$L$; see Eqs.~(\ref{karma}, \ref{karma0}). It may be useful to repeat
that the most unusual aspect of this phase-transition is that its
point is strictly realization-dependent.  The same philosophy can now
be applied to Figs.~(\ref{fig_3-11a}, \ref{fig_3-2a}): there are
realization-dependent values of $f$, where $\x$ has jumps of order of
$L/2$ (recall that for the figures we have taken $L=10^4$ or
$L=5\times 10^4$). It is seen as well that there can be several such
phase-transitions for a single system (single realization of noise).
The latter fact can by itself appear to be rather surprising. However,
it is known that some disordered \cite{de}, or deterministic but
strongly-frustrated \cite{saqo}, systems can experience several phase
transitions; there can even exist quasi-continuous domains of
criticality \cite{saqo}.  With the same logic one sees that the
typical realizations presented in Figs.~(\ref{fig_3-1b},
\ref{fig_3-2b}) do not have phase transitions in the domain $0<f<1$ at
least for the considered values of $L$.

\subsection{The behavior of $\langle \x\rangle$.}

As we already noted, once the effect of non-self-averaging is present,
the basic physical quantities are the typical realizations, since it
is these features that are directly observed in experiments.  It is,
however, still of relevance to know the behavior of the average number
of unzipped bonds $\langle\x\rangle$, since it illustrates what are
the precise differences as compared to typical realizations.

Here we report on two features of $\langle\x\rangle$ as a function of
$f$.  The first one is how does $\langle\x\rangle$ depend on $f$ for
small values of $f$. In particular, is there any power law dependence
similar to $\langle\x(f)\rangle \propto f^{-2}$ present in the uncorrelated
noise situation, and verified by us numerically in section \ref{un}?
Note that for the long-range correlated situation with the index $\alpha$
such a power law 
\BEA
\label{hreaner}
\langle\x\rangle\propto f^{-2/\alpha}, \EEA 
was recently predicted for small $f$; see Ref.~\cite{nelson}.  The
most adequate way to look for the power-laws is to plot
$-\ln\langle\x\rangle$ as a function of $\ln f$, then a power-law
should display itself via a straight line.  Fig.~(\ref{fig_6-2a})
display such a plot obtained for various values of $\alpha$ and
$L=5\times 10^4$.  The quantity $\langle \x\rangle$ was calculated by
direct averaging over $10^3$ realizations and the results were checked
for stability upon increasing (by two times) the number of
realizations.  As seen, this figure shows very weak dependence of
$-\ln\langle \x\rangle$ on $\ln f$. There are no 
convincing indications of a power law. In particular, when decreasing
$\alpha$ the dependence of $-\ln\langle \x\rangle$ on $\ln f$ does
become {\it weaker}, in obvious contrast with the prediction made by
Eq.~(\ref{hreaner}).  For $\alpha\to 0$ this behavior coincides with
those of the exact solution discussed in section \ref{total}.

It should be noted that $-\ln\langle \x\rangle$ is a perfectly smooth
function of $\ln f$: all jumps and flat regions present for the first
class of typical realizations | which involves the majority of
realizations | became washed out when averaging over $10^3$
realizations. This gives another indication that the point of jumps in
the above class are completely random and vary from one realization to
another.

Once we realized that in a rather wide interval of $f$'s
| typically $\ln f<-0.5$, as seen in Fig.~(\ref{fig_6-2a}) |
the dependence of $\langle\x\rangle$ on $f$ is weak, 
we have studied the behavior of
$\langle\x(f=0)\rangle$ as a function of $L$ and $\alpha$. As shown by
Figs.~(\ref{fig_7-2b}, \ref{fig_7-3b}, \ref{fig_7-4b}) 
numerical results fit well into the following scaling equation:
\BEA
\label{tiratsu}
\langle\x(f\to 0)\rangle\propto\left(\frac{L}{2}\right)^{1+\delta (\alpha)}.
\EEA
The values of $\delta(\alpha)$ for several $\alpha$'s are shown by
Fig.~(\ref{fig_8}). For $\alpha\approx 0$, we get $\delta=0.01$ which
is in a good agreement with exact value $\delta(\alpha=0)=0$ got in section
\ref{total}.

Two important features of result (\ref{tiratsu}) are to be mentioned.
First, as seen from Figs.~(\ref{fig_7-2b}, \ref{fig_7-3b},
\ref{fig_7-4b}) the value of $\langle\x\rangle(f=0)$ adequately
characterizes the whole domain of small $f$, since the dependence of
$\langle \x\rangle $ on $f$ is weak.  Second, as seen from
Fig.~(\ref{fig_8}), the function $\delta(\alpha)$ increases with
$\alpha$, but saturates for $\alpha\geq 0.5$ at $\delta=0.08$.  It
appears that the same result (\ref{tiratsu}) with the index
$\delta=0.08$ holds for the uncorrelated noise, but there its region
of validity is restricted (for $L=5\times 10^4$, $T=\Delta=1$) by very
small $f<0.01$ values of $f$, in contrast to the long-range
situation. Thus, as far as the small-$f$ characteristics are concerned the
result (\ref{tiratsu}) seems to be universal, and it is likely that
$\delta(\alpha)$ can have the same status as critical indices in the
usual theory of phase transitions.

We conclude by repeating two main qualitative features of the
average number $\langle\x(f)\rangle$ of unzipped bonds as revealed by
our numerical analysis: in the long-range situation and for small
forces $f$, the behavior of $\langle\x(f)\rangle$ as a function of $f$
does not display any power-law, and is governed by its value at
$f=0$. The latter one satisfies to power-law (\ref{tiratsu}) as a
function of $L$.

\section{Summary and conclusion.} 

In the present paper we have studied how statistical correlations
present in the base-sequence of a DNA molecule influence the process
of unzipping.  There were two related motivations for our study. On
the one hand, the existence of these correlations | that can have both
finite-range and long-range character | is by now a well-established
fact \cite{longrange,voss,coding,review}.  It is, therefore,
legitimate to study how they influence on the DNA physics.  On the
other hand, general qualitative predicitions drawn on the above
influence can be used for explaininig the reason of rich correlated
structures found in the base-sequence of DNA. Recall that various
segments of a DNA molecule can have different | finite-range or
long-range \cite{coding,review} | correlation structures.  Moreover,
DNA molecules belonging to different evolutionary classes have
different correlation properties of their base-sequences \cite{voss}.

The model we studied contains only the most minimal number of
ingredients needed to describe unzipping, and to account for
correlations in the base-sequence of DNA. Therefore, many realisitic features
of the unzipping process remain beyond of our study. We, nevertheless,
believe that the obtained results will be useful especially for drawing
qualitative conclusions.

Let us now shortly summarize our results starting from the
finite-range correlated situation. In section \ref{OUnoise} we have
shown that the presence of a finite correlation length $\tau$ plays a
stabilizing role for the unziping process: for a fixed external force
$f$ the average number of broken base-pairs decreases under increasing
of $\tau$. If only finite-range correlations are present, the process
of unzipping does not depend much on the detailed structure of the
base-sequence: all typical | i.e., frequently met among all possible
base-sequeces | scenarios of unzipping have the same qualitative
pattern of behavior, that is, the number of the broken base-pairs
diverges as the external force approaches its critical value: $f\to
0$. This divergence can be adequately understood by studying the
average | over all possible base-sequences | number of broken
base-pairs.  All by all, one can say that the basic influence of
finite-range correlations is in stabilizing the DNA molecule with
respect to the external unzipping force.

The influence of long-range correlations is certainly more drastical.
Possibly the most important aspect is that the situation is
essentially non-self-averaging: there are two radically different
scenarios of typical unzipping which depend on the detailed structure
of the base-sequence and which do not coincide with the behavior
averaged over all possible base-sequences.  Within the first scenario,
the number of broken base-pairs $X(f)$ shows as a function of the
external force $f$ a sequence of sharp jumps at sequence-dependent
values of $f$. The overall number of jumps is nearly constant within
the class. Each jump has the magnitude comparable with $L$, that is,
under small change of $f$ a large number of base-pairs can be opened.
The point $f=0$ is special, since $X(0)$ either coincides with $L$, or
at least is very close to it. We argued in section \ref{typical} that
it is sensible to describe this scenario as a sequence of
phase-transitions. Such an effect is known from other disordered or
strongly-frustrated systems \cite{de,saqo}.

The second typical scenario is crucially different.  Now $X$ is a
smooth, slowly changing function of the external force $f$ in the
whole relevant domain $0<f<1$. There is no any sign of phase
transition, and the value $f=0$ is not distinguished from $f>0$ as far
as $X$ is concerned.  DNA molecules which due to the structure of their
base-sequence fall into this class are thus rather stable with respect
to the external unzipping force.

It appears, interestingly enough, that the qualitative and even some
quantitative feautures of the long-range correlated situation can be
understood via the analytical solution of the model with the totally
correlated (frozen) noise, which we presented in section \ref{total}.
In particular, this allows to explain why there exist two typical
scenarios with widely different behavior of the number of unzipped
base-pairs, and provides rather robust analytical indications for the
phenomenon of non-self-averaging.

Summarizing features of these two scenarios, one can say that
long-range correlations increase the adaptability of the corresponding
DNA molecule, since in some typical scenarios it becomes more stable
with respect to the force (any sharp transition is absent), while in
others the unzipping is realized via a sequence of sharp phase
transitions. The actual scenario for a single molecule will crucially
depend on the detailed structure of the base-sequence.

We also studied how the average number $\langle\x\rangle$ of unzipped
base-pairs depends on the applied force $f$. In contrast to
white-noise situation, where the behavior of $\langle\x\rangle$ for
small (that is, critical) forces $f\to 0$ is governed by a power-law $\langle
\x\rangle\sim f^{-2}$, we found numerically no indications of a
power-law for small forces in the long-range correlated situation.  In
contrast, the dependence of $\langle \x(f)\rangle$ on $f$ for small
$f$'s is very weak and to a large extent is governed by
$\langle\x(f=0)\rangle$. The latter quantity displays a power-law
behavior (\ref{tiratsu}) as a function of $L$. The region of validity
of this power-law appeared to be unexpectedly wide.

We hope that these results will contribute into understanding of the
role and the purpose of correlations structures in DNA.

\section*{Acknowledgments}
This work of Zh.S. Gevorkian, 
C.-K. Hu and W.-C. Wu was supported in part by a grant 
from National Science Council in Taiwan under Grant NSC 92-2112-M-001-063.  

The work of A.E. Allahverdyan is part of the research programme of the
Stichting voor Fundamenteel Onderzoek der Materie (FOM, financially
supported by the Nederlandse Organisatie voor Wetenschappelijk
Onderzoek (NWO)).
 
Zh.S. Gevorkian acknowledges interesting discussions with A. Maritan
and D. Marenduzzo.

\appendix

\section{Generation of the long-range correlated noise.}
\label{num}

Using ideas of \cite{sancho} we shall here describe a method for
numerical generation of a Gaussian random noise $\eta(t)$ with zero
average and an arbitrary symmetric autocorrelation function:
\BEA
\label{burb}
K(t-t')=\langle\, \eta(t)\,\eta(t')\,\rangle,
\EEA
\BEA
\label{barb3}
K(t)=K(-t). 
\EEA
Assume that the noise is periodic with period $M$:
\BEA
\eta(t)=\eta(t+M).
\EEA
Therefore $K(t)$ is also periodic with the same period and can be
expanded as
\BEA
\label{barb1}
K(t)=\sum_{n=-\infty}^{\infty}k_n\,e^{-in\omega_0t},\qquad
\omega_0=\frac{2\pi}{M},
\EEA
where $k_n$ is given by Fourier formula:
\BEA
\label{barb2}
k_n=\frac{1}{M}\int_{-M/2}^{M/2}\d t\, K(t)\,e^{in\omega_0t}.
\EEA
Since $K(t)$ is a real and symmetric function, 
$k_n=k^*_{n}=k_{-n}$, and thus
\BEA
k_n=\frac{2}{M}\int_{0}^{M/2}\d t\, K(t)\,\cos(n\omega_0t).
\label{bobo}
\EEA

It is now straightforward to see that the noise $\eta$ we are looking
for is represented as
\BEA
\eta=\sum_{n=-\infty}^{\infty}\sqrt{k_n}\,\eta_n\,
e^{-in\omega_0t},
\EEA
where $\eta_n$ are complex Gaussian random variables with 
\BEA
\langle\,\eta_n\eta_m\,\rangle=\delta(n+m),
\EEA
where $\delta(0)=1$ and $\delta(k)=0$ for $k\not =0$. 
Indeed, once $\eta_n$ are assumed to be Gaussian, $\eta(t)$
is Gaussian as well; it is seen as well that (\ref{burb}) is valid.
Complex random variables $\eta_n$ can be conveniently expressed via
real random variables:
\BEA
&&\eta_n=\frac{1}{\sqrt{2}}\left(a_n+ib_n\right),\qquad 
{\rm for}\qquad n\geq 1,\\
&&\eta_n=\frac{1}{\sqrt{2}}\left(a_n-ib_n\right),\qquad 
{\rm for}\qquad n\leq -1,\\
&&\eta_0=a_0,
\EEA
where are quantities $a_n$ and $b_n$ are independent,
zero-average Gaussian random variables normalized to one:
\BEA
\langle a_ka_l\rangle=\delta_{kl},\qquad
\langle b_kb_l\rangle=\delta_{kl},\qquad
\langle a_kb_l\rangle=0.
\EEA
Using this one writes
\BEA
\label{gg1}
\eta=\sum_{n=1}^{\infty}\sqrt{2k_n}[\,a_n\cos(n\omega_0t)+b_n
\sin(n\omega_0t)\,]+\sqrt{k_0}\,a_0.
\EEA

Let us consider an example:
\BEA
K(t)=&&\sigma,\qquad {\rm for}\qquad t<1,\nonumber\\
=&&\frac{\sigma}{\sqrt{t}},\qquad {\rm for}\qquad t\geq 1.
\EEA
This represents a long-range correlated noise regularized
for small $t$. For this autocorrelation function the coefficients
$k_n$ read from (\ref{bobo}):
\BEA
\label{gg2}
k_0=2\sigma\left(
\frac{\sqrt{2}}{\sqrt{M}}-\frac{1}{M}
\right),\qquad
k_n=\frac{\sigma \sin(n\omega_0)}{\pi n}+
\frac{2\sigma}{\sqrt{n \,M}}\left[
F_C\left(\sqrt{2n}\right)-F_C\left(\frac{\sqrt{2n}}{\sqrt{M/2}}\right)
\right],
\EEA
where $F_C(x)$ is Fresnel's $C$-function:
\BEA
F_C(x)=\int_0^x\d t\,\cos\left(
\frac{\pi t^2}{2}\right).
\EEA

\subsection{Numerical implementation.}

Eqs.~(\ref{gg1}, \ref{gg2}) are sufficient for generating long-range
correlated, periodic Gaussian random noise. However, for numerical
implementations this noise has to be discretized. First we note that
the above method produces periodic random noise (with period $M$),
while our problem does not have any periodicity. Therefore, we have
chosen $M=2L$, and took discrete values of $t=1,2,...,L$ in
Eq.~(\ref{gg1}), thereby generating $L$ long-range correlated random 
numbers without any periodicity. Note that (\ref{gg1}) contains
infinity as the upper limit of the summation in its RHS. For numerics 
this infinity should obviously be substituted by some number larger 
than $L$, and additionally one should check that the situation is
stable with respect of varying this number. As for concrete
calculations we have used, e.g., $L=10^4$, we found sufficient to take
for this upper summation limit $10^4$.

Numerical simulations in section \ref{typical} were performed by using the
gaussian independent random variables generated by the "gasdev" algoritm of
Ref.~\cite{27}. The long-range correlated noise was
generated following the scheme proposed in Appendix A.

\section{Derivation of two asymptotic relations.}
\label{aux}

Here we derive the following asymptotic identities used in the main
text:
\BEA
\label{kont}
\int_{a}^\infty\frac{\d\xi}{\sqrt{2\pi}}\,e^{-\xi^2/2}=
\frac{e^{-a^2/2}}{a\sqrt{2\pi}}\,\left(1-\frac{1}{a^2}+\dots\right),
\qquad a\gg 1,
\EEA
\BEA
\int_0^{a}{\d\xi}\,e^{\xi^2/2}=
\int_0^{a}{\d\xi}\,e^{\xi^2/2}=
\frac{e^{a^2/2}}{a}\left(1+\frac{1}{a^2}\right)+
\dots,\qquad a\gg 1.
\label{konto}
\EEA

The first one is easliy done via integration by parts:
\BEA
\int_{a}^\infty\frac{\d\xi}{\sqrt{2\pi}}\,e^{-\xi^2/2}=
-\int_{a}^\infty\frac{\d[\,e^{-\xi^2/2}\,]}{\xi\,\sqrt{2\pi}}\,
=\frac{e^{-a^2/2}}{a\sqrt{2\pi}}+
\int_{a}^\infty\frac{\d[\,e^{-\xi^2/2}\,]}{\xi^3\,\sqrt{2\pi}}.
\EEA

For the second relation 
one notes that for $a\gg 1$ the relevant domain of integration is
$\xi \sim a$. In more details,
\BEA
\int_0^{a}{\d\xi}\,e^{\xi^2/2}=
e^{a^2/2}\,\int_0^{a}{\d\xi}\,e^{-a\xi+\xi^2/2}=
\frac{e^{a^2/2}}{a}\,
\int_0^{a^2}{\d y}\,e^{-y+y^2/(2a^2)}.
\label{kont1}
\EEA
Now one can expand inside of the second exponent in the RHS of 
(\ref{kont1}), since the main contribution to the integral comes
from $y\sim 0$ (the other side, that is $y\sim a^2$, is strongly 
suppressed as seen):
\BEA
\frac{e^{a^2/2}}{a}\,
\int_0^{a^2}\d y\,e^{-ay+y^2/(2a^2)}
=\frac{e^{a^2/2}}{a}\,
\int_0^{a^2}\d y\,e^{-y}\left(1+\frac{y^2}{2a^2}
+\dots\right).
\EEA
Neglecting exponentially small terms, one gets get finally (\ref{konto}).


\begin{thebibliography}{99}


\bibitem{bio} 
D. Freifelder and G.M. Malacinski, {\it Essentials of Molecular
  Biology}, (Jones and Bartlett Publishers, Boston \& London, 1993). 

\bibitem{biofiz}
A.Yu. Grosberg and A.R. Khokhlov, {\it Statistical Physics of 
Macromolecules}, (American Institute of Physics, New-York, 1994).


\bibitem{hindu} S.M. Bhattacharjee, J. Phys. A, {\bf 33}, L423 (2000).
K.L. Sebastian, Phys. Rev. E, {\bf 62}, 1128 (2000).
E. Mukamel and E. Shakhnovich, cond-mat/0108447. 
E. Orlandini, et al., cond-mat/0109521.
S. Cocco, et al., 
cond-mat/0206238.

\bibitem{non-hindu}
S.M. Bhattacharjee and D. Marenduzzo, J. Phys. A, {\bf 35}, L141
(2002); cond-mat/0106110.

\bibitem{maritan}
D. Marenduzzo, A. Trovato and A. Maritan, Phys. Rev. E, {\bf 64},
031901 (2001); Phys. Rev. Lett., {\bf 88}, 028102 (2002).

\bibitem{nelson}D.K. Lubensky and D.R. Nelson, Phys. Rev. Lett.,
{\bf 85}, 1572 (2000); Phys. Rev. E, {\bf 65}, 031917 (2002).

\bibitem{exp}B. Essevaz-Roulet, 
U. Bockelmann and F. Heslot,
Proc. Nat. Acad. Sci., {\bf 94}, 11935 (1997); 
T.T. Perkins, D.E. Smith and S.Chu, 
Science, {\bf 264}, 819 (1994). R. Merkel, et
al., Nature, {\bf 397}, 50 (1999).

\bibitem{expreview}T.R. Stick, et al., Rep. Prog. Phys., {\bf 66},
1 (2003).

\bibitem{longrange}W. Li, Int. J. Bif \& Chaos, {\bf 2}, 137 (1992).
I. Amato, Science, {\bf 257}, 74 (1992). W. Li and
K. Kaneko, Eur. Phys. Lett. {\bf 17}, 655 (1992). 
C.K. Peng, et al., Nature, {\bf 356}, 168 (1992).
S. Buldyrev, et al., Phys. Rev. E, {\bf 47}, 4514 (1993); {\it ibid},
{\bf 51}, 5084 (1995). 

\bibitem{voss}R. Voss, Phys. Rev. Lett. {\bf 68}, 3805 (1992).
X. Lu, et al., Phys. Rev. E, {\bf 58}, 3578 (1998).


\bibitem{coding}C.A. Chatzidimitrou-Dreismann and D. Larhammar, Nature
{\bf 361}, 212 (1993). V.V. Prabhu and J.M. Claverie, Nature {\bf
359}, 782 (1992).  A.K. Mohanti and A.V. Narayana Rao,
Phys. Rev. Lett.  {\bf 84}, 1832 (2000). B. Audit, et al.,
Phys. Rev. Lett.  {\bf 86}, 2471 (2001). S. Guharay, et al., Physica D
{\bf 146}, 388 (2000).

\bibitem{review}W. Li, Comput. Chem. {\bf 21}, 257 (1997).

\bibitem{maria}M. Vieira, Phys. Rev. E {\bf 60}, 5932 (1999).

\bibitem{jenya}E.S. Mamasakhlisov, et al., 
J. Phys. A {\bf 30}, 7765 (1997).


\bibitem{manfred}M. Opper, J. Phys. A {\bf 26}, L719 (1993).
C. Monthus and A. Comtet, J. Phys. I (France) {\bf
4}, 635 (1994). 

\bibitem{azbel}M.Ya. Azbel, Phys. Rev. Lett, {\bf 31}, 589 (1973).

\bibitem{risken}H. Risken, {\it The Fokker-Planck Equation}, 
(Springer-Verlag, Berlin, 1989).

\bibitem{hanggi}P. Jung and P. H\"anggi, Phys. Rev. A {\bf 35}, 4467 (1987)

\bibitem{fox} R. Fox, Phys. Rev. A {\bf 33}, 467 (1986)

\bibitem{sancho}A.H. Romero and J.M. Sancho, J. Comp. Phys., {\bf
156}, 1 (1999); cond-mat/9903267. H.A. Makse, et al., Phys. Rev. E, 
{\bf 53}, 5445 (1996).

\bibitem{f1}Note, however, that there are exclusions from this
rule for certain bacteria \cite{bio}; for them the concentration of GC
pairs can differ substantially from that of AT pairs. These situations
are, nevertheless, fairly rare.

\bibitem{gross}D.H.E. Gross, {\it Microcanonical Thermodynamics: Phase
Transitions in Finite Systems}, (World Scientific, Singapure, 2001).

\bibitem{brout}R. Brout, Phys. Rev. {\bf 115}, 824 (1959).

\bibitem{de}B. Derrida, Phys. Rep., {\bf 103}, 29 (1984).

\bibitem{saqo}
A. E. Allahverdyan, N. S. Ananikian, and S. K. Dallakian,
Phys. Rev. E, {\bf 57}, 2452 (1998). 


\bibitem{27} W.H.Press, S.A.Tenkolsky, W.T.Vetterling and B.P.Flanneri,
{\it Numerical Recipes in Fortran: The Art of scientific Computing}, 2nd
edn.(Cambridge University Press, USA, 1992).

\end{thebibliography}
\end{document}